\documentclass[aps,twocolumn,showpacs,superscriptaddress]{revtex4}

\usepackage[dvips]{graphicx}
\usepackage{amsmath}
\usepackage{amsfonts}
\usepackage{amssymb} 

\newcommand{\vecb}[1]{{\bf{#1}}}
\newcommand{\cvec}{\vecb c}

\newcommand{\ul}{\underline}

\newcommand{\td}{\tilde}

\newcommand{\Tr}{\mathop{\mathrm{Tr}}}

\newcommand{\upd}{{\mathrm{d}}}

\newcommand{\gmat}[1]{\boldsymbol{#1}}
\newcommand{\gmatx}[1]{\td{\gmat{#1}}}

\newcommand{\op}{\hat} 
\newcommand{\im}{\mathop{\textrm{Im}}}
\newcommand{\re}{\mathop{\textrm{Re}}}

\newcommand{\iu}{{\mathrm{i}}}

\begin{document}

\title{Electron-vibration interaction in transport through atomic
gold wires} 

\author{J. K. Viljas}
\affiliation{Institut f\"ur Theoretische Festk\"orperphysik,
Universit\"at Karlsruhe, 76128 Karlsruhe, Germany}

\author{J. C. Cuevas}
\affiliation{Institut f\"ur Theoretische Festk\"orperphysik,
Universit\"at Karlsruhe, 76128 Karlsruhe, Germany}
\affiliation{Departamento de F\'{\i}sica Te\'orica de la Materia
Condensada, Universidad Aut\'onoma de Madrid, 28049-Madrid, Spain}

\author{F. Pauly}
\affiliation{Institut f\"ur Theoretische Festk\"orperphysik,
Universit\"at Karlsruhe, 76128 Karlsruhe, Germany}

\author{M. H\"afner}
\affiliation{Institut f\"ur Theoretische Festk\"orperphysik,
Universit\"at Karlsruhe, 76128 Karlsruhe, Germany}

\date{\today}

\begin{abstract}
We calculate the effect of electron-vibration coupling
on conduction through atomic gold wires, which was 
measured in the experiments of Agra\"it \emph{et al.} 
[Phys.\ Rev.\ Lett.\ \textbf{88}, 216803 (2002)]. 
The vibrational modes, 
the coupling constants, and the inelastic
transport are all calculated using a tight-binding parametrization
and the non-equilibrium Green function formalism.
The electron-vibration coupling gives
rise to small drops in the conductance at voltages
corresponding to energies of some of the vibrational modes.
We study systematically how the position and height of these
steps vary as a linear wire is stretched and more atoms are added 
to it, and find a good agreement with the experiments. 
We also consider two different types of geometries, which 
are found to yield qualitatively similar results.
In contrast to previous calculations, we find that
typically there are several close-lying drops due 
to different longitudinal modes.
In the experiments, only a single drop is usually visible,
but its width is too large to be accounted for by temperature.
Therefore, to explain the experimental results, 
we find it 
necessary to introduce a finite
broadening to the vibrational modes, which makes the 
separate drops merge into a single, wide one.
In addition, we predict how the signatures of 
vibrational modes in the conductance curves 
differ between linear and zigzag-type wires.

\end{abstract}


\maketitle

\section{introduction}

The electron transport properties of atomic point contacts between two
metallic electrodes have been intensively studied during the past decade 
\cite{Review03}. Contacts of this type are typically formed by using 
mechanically controllable break junctions (MCBJ) or with the tip of a 
scanning tunneling microscope (STM). It has been found that the conductance 
of such contacts depends strongly on the electronic structure of the metals,
and for monovalent metals there is a tendency for quantization in units of 
the conductance quantum $G_0=2e^2/h$ \cite{ScheerAgrait98}. Point contacts 
formed from  Gold (Au), Platinum (Pt), or Iridium (Ir) by one of the MCBJ 
or STM methods have the further interesting property of being able to 
sustain single-atom thick chains, so called atomic wires 
\cite{OhnishiKondoTakayanagi98,Yanson98,SmitUntiedt01}.
Following their discovery, a good amount of experimental 
\cite{UntiedtYanson02,SmitUntiedt03} and theoretical 
\cite{Emberly99,BrandbygeMozos02,Palacios02,Zhuang04,LeeBrandbyge04,Vega04} 
work has been carried out to further investigate the conduction properties
of atomic wires. It is by now well established that the zero-bias conductance
of gold wires is close to one $G_0$ due to a single, almost fully open 
transmission channel at the Fermi energy 
\cite{ScheerAgrait98,Ludoph99,Scheer01,Rubio03}.  However, less detailed
work has been done in the study of truly nonequilibrium properties, such as
the current-voltage characteristics \cite{NielsenBrandbyge02}.

In recent experiments, the conductance vs. voltage characteristics $G(V)$ of
gold wires formed by the STM technique were measured \cite{AgraitUntiedt02}.
It was observed that the conductance often has a very pronounced, single drop
from $G_0$ at a critical voltage $V_{ph}=10-20$ mV, marking the onset of a 
dissipative process. The size of the drop was on the order of $0.5\%-2.0\%$ 
of $G_0$. It was also found that stretching of the wire typically leads to 
an increase in the step, and to a decrease in the critical voltage $V_{ph}$.
Based on simple arguments for infinite single-orbital tight-binding chains,
these findings were interpreted as a sign of the excitation of
vibrational modes in the wire: only a single longitudinal mode with
twice the Fermi wave vector can be excited, since this corresponds to the 
momentum which must be transferred from an electron to the vibrations
in a single backscattering process. Although the validity of such arguments
for a wire of finite length (of typically less than 10 atoms) can be 
questioned, the interpretation was backed up by first-principles 
calculations \cite{Frederiksen04a,Frederiksen04b}.
The authors of Ref.\ \onlinecite{Frederiksen04a}
emphasize the importance of so-called alternating bond length (ABL)
modes, and in particular the longitudinal mode of highest frequency.

Although it seems evident that the interpretation based on vibrational 
modes is essentially correct, what is still lacking is a systematic 
study of the behavior of wires with varying numbers of atoms, and
surrounded by various lead geometries. Many questions of the basic physics 
are also still not very well understood: When exactly does the 
electron-vibration coupling lead to a drop and when an increase 
in the conductance? Why does there appear to be just a single
drop in the experiments of Ref.\ \onlinecite{AgraitUntiedt02}, although
the momentum conservation is not exact? 
What determines the height and width of this drop? Below we aim to 
discuss the possible answers to some of these questions.

In this paper we concentrate on studying the current-voltage 
characteristics of gold wires. 
We use a Slater-Koster \cite{SlaterKoster54} type tight-binding (TB)
approach, where the parameters are taken from the non-orthogonal 
parametrization of Papaconstantopoulos and coworkers 
\cite{CohenMehlPapa94,MehlPapa96,PapaMehl03,website}. 
The use of such a parametrization 
\cite{MontgomeryHoekstra03,Emberly00,BrandbygeKobayashiTsukada99} 
makes the modeling of atomic wires
computationally less intensive as compared with fully \emph{ab initio}
methods. 
The approach is still microscopic in that it takes
into account the symmetries of the atomic $s$, $p$, and $d$ 
valence orbitals, which, via hybridization, form the conduction channels.
It is also general enough to allow one to model
everything within the same framework: we use the parameters to compute the 
total energy of the wire, and thus optimize the geometry. After this, 
the normal modes of oscillation and the electron-vibration coupling 
constants can be computed. Finally, we calculate the transport properties
using the non-equilibrium Green function (or Keldysh) approach.
Our implementation is very similar to that of 
Ref.\ \onlinecite{BrandbygeKobayashiTsukada99}, and
the present work is, in essence, an extension of that to inelastic 
transport.
In addition to the full \emph{ab initio} calculations \cite{Frederiksen04a},
the effect of electron-vibration interactions on transport through 
molecular wires has been recently studied by some simple single-level
models \cite{MiiTikhodeevUeba03,MitraAleinerMillis04,GalperinRatnerNitzan04}.
The tight-binding approach stands somewhere in between these two 
extremes.

We compute the conductance to
lowest nontrivial order in the electron-vibration 
coupling constant. There are essentially two well-defined limits
which can be studied. In the first limit the vibrational-mode distribution 
remains in equilibrium due to a strong coupling of the modes to 
an external equilibrium bath formed by the leads.
In the opposite limit the distribution 
is driven to strong non-equilibrium by the bias voltage.
These are the \emph{externally damped} and the 
\emph{externally undamped} limits of 
Ref.\ \onlinecite{Frederiksen04a}.
However, in the first limit one should also account for the 
strong broadening of the vibrational modes.
We derive equations which take this into account 
in a phenomenological manner.

With our simple, self-contained method of optimizing the
geometry, we obtain vibrational frequencies which are of 
the correct order of magnitude, usually to within a factor 
of two. 
We study how the positions and heights of the conductance drops due to the 
electron-vibration coupling vary with elongation of a linear wire,
and find a good overall agreement with the experiments
of Ref.\ \onlinecite{AgraitUntiedt02}.
The height of the conductance steps grows together with the 
length of the wire, being typically of the order 
0.5\% --- 5\% of $G_0$ for wires of 11 atoms or less.
As found in the earlier calculations \cite{Frederiksen04a}, we 
find that the highest-frequency longitudinal modes usually couple most 
strongly, although there seems to be no fundamental reason 
for a bias toward the ``ABL'' modes.
But in contrast to previous theoretical results, 
the conductance drop is usually found to 
occur in two or more consecutive steps which are due to several
close-lying longitudinal vibrational modes. 
Thus we find the ``mode selectivity'' to be only very approximate.
However, the steps can be made to merge into a single
one, when we introduce a large enough phenomenological 
broadening to the vibrational modes, such that the experimentally
observed step widths of $\sim 5$ meV are accounted for.
We also briefly study the conductance signatures of chains
which have a zigzag-like form, instead of the linear one.

The paper is divided into the following parts. 
In Sec.\ \ref{s.def} we start by defining the problem,
and discussing the electron-vibration 
coupling. 
In Sec.\ \ref{s.methods} we briefly discuss the 
calculation of the vibrational modes and the 
electron-vibration coupling constants,
as well as our methods of
computing the transport.
After this, Sec.\ \ref{s.wbl} introduces the important 
wide-band approximation to the full formalism.
In Sec.\ \ref{s.simple} we use the formalism to 
analyze simple tight-binding models, and
in Sec.\ \ref{s.spd} the full $spd$-tight-binding model.
Section \ref{s.conc} ends with some conclusions and discussion. 
Technical details which are not of immediate importance
are postponed to the appendixes. These include a
discussion of the calculation of the matrix elements needed for the
coupling constants.
Readers mainly interested in the results can directly
jump to Secs.\ \ref{s.simple}--\ref{s.conc}.

\section{Definition of the problem} \label{s.def}

\begin{figure}[!tb] 
\includegraphics[width=0.9\linewidth,clip=]{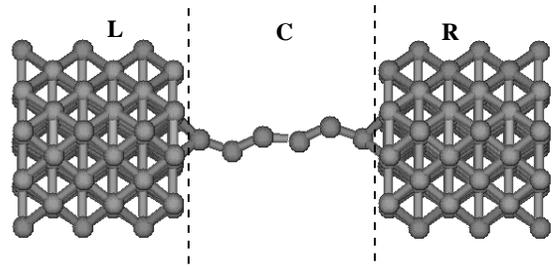}
\caption{Geometry A, without ``pyramids''. 
A zigzag wire with $N_{ch}=6$ atoms is shown.} 
\label{f.lcr}
\end{figure}
\begin{figure}[!tb] 
\includegraphics[width=0.9\linewidth,clip=]{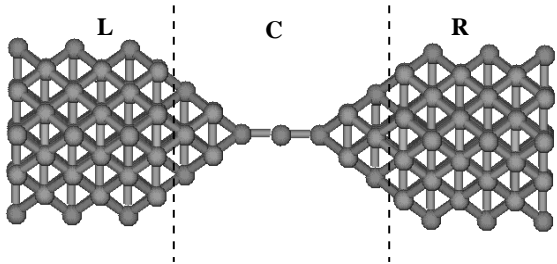}
\caption{Geometry B, with ``pyramids'' and a linear wire of 
$N_{ch}=3$ atoms.} 
\label{f.lcr_pyra}
\end{figure}

To model transport through atomic wires, we consider two idealized geometries, 
shown in Figs.\ \ref{f.lcr} and \ref{f.lcr_pyra}. We call these geometry 
A and geometry B, respectively. 
Both involve a gold chain of $N_{ch}$ atoms suspended between 
two gold leads. 
As the leads, we simply use semi-infinite ``bars'', where the repeat unit consists 
of two layers, with 12 and 13 atoms respectively, mimicking an infinite fcc [001] 
surface, where the $z$ axis is always chosen parallel to the axis of the wire. 
The particular choice for the leads should not be very important, as long as 
they are infinite in one direction, and wider than the contact region.
In geometry B, the chain connects to
small clusters of atoms or
``pyramids'' on the surfaces (in our case consisting of 9 atoms), making it 
perhaps the more realistic one of the two. For technical reasons, the geometry is 
divided into three parts, the semi-infinite left ($L$) and right ($R$) leads, 
and the ``central cluster'' ($C$), which also includes the pyramids if any.
These parts are indicated in the figures.
We shall also consider some simple models where the regions $L$, $C$, and $R$ 
are parts of an infinite linear wire. 

Our objective is to model the effect of vibrations (or ``phonons'') of the wire
on the transport, when a voltage is applied over the contact.
Within a tight-binding picture, the system of electrons coupled to vibrational 
modes is described by the 
Hamiltonian
\begin{equation}
\op{H} = \op{H}_{e} + \op{H}_{vib} + \op{H}_{e-vib} , \nonumber
\end{equation}
where
\begin{equation}\begin{split} \label{e.ham1}
\op{H}_{e}&=\sum_{ij}{d}^\dagger_iH_{ij}{d}_j \\
\op{H}_{vib}&=\sum_{\alpha}\hbar\omega_\alpha{b}^\dagger_\alpha{b}_\alpha \\
\op{H}_{e-vib}&=
\sum_{ij}\sum_{\alpha}
{d}_i^\dagger \lambda_{ij}^\alpha {d}_j
({b}_\alpha^\dagger+{b}_\alpha) .
\end{split}\end{equation}
Here $\omega_\alpha$ are the vibrational frequencies,
$H_{ij}=\langle i|H_{}|j\rangle$ are the 
matrix elements of the equilibrium single-electron
Hamiltonian $H_{}$ in the atomic-orbital basis $\{|i\rangle\}$,
and $\lambda_{ij}^\alpha$
are the electron-vibration coupling constants.
The index $i$ denotes collectively the atomic sites and orbitals, and
$\alpha$ runs from $1$ to $3N_{vib}$, where
$N_{vib}$ is the number of atoms in the system, which are allowed to vibrate.
The creation and annihilation operators 
for vibrational modes $b_\alpha^\dagger$,$b_\alpha$ satisfy
the bosonic commutation relation
$[{b}_\alpha,{b}_\beta^\dagger]=\delta_{\alpha\beta}$.
The electronic basis is in general non-orthogonal, with
overlap matrix elements $S_{ij}=\langle i|j\rangle$.
Thus the anticommutator for electron operators $d^\dagger_i$,$d_i$
is given by $\{d_i,d^\dagger_j\}=[\gmat{S}^{-1}]_{ij}$.

Hereafter we denote the matrices with components
$A_{ij}$ with a boldface symbol $\gmat{A}$.
The matrices $\gmat{H}$, $\gmat{S}$, and $\gmat{\lambda}^\alpha$
are all symmetric in our case.
In the $spd$ TB model, the matrix elements $H_{ij}$ and $S_{ij}$ are
obtained directly from the parametrization \cite{website}.
These can also be used to calculate the
the vibrational frequencies $\omega_\alpha$ and the coupling constants 
$\gmat{\lambda}^\alpha$, as we shall now describe.

\section{Methods} \label{s.methods}

The solution of the inelastic transport problem 
involves a few rather separated sub-problems: the
optimization of the geometry and 
evaluation of the vibrational modes, estimation of the 
electron-vibration coupling constants, and finally the calculation 
of the transport.
In the following we give only a brief description of each of them,
and refer the reader to appendixes for details.
Our basic approach is to solve for the elastic transmission 
problem exactly, and then to take the electron-vibration
coupling into account in a slightly modified
version of lowest-order perturbation theory.
Other works have considered the so-called self-consistent Born 
approximation (SCBA) \cite{Frederiksen04a,GalperinRatnerNitzan04},
where some of the terms in the perturbation expansions are 
effectively summed to infinite order.
However, this is not essential for describing the
basic physics which is involved in the present problem.

\subsection{Vibrational modes and the electron-vibration coupling constants}

The calculation of the vibrational modes requires 
knowledge of the total (ground-state) energy of the system as
a function $E(\vec R_k)$ of the ionic coordinates 
$\vec R_k$ with $k=1,\ldots,N_{vib}$.
This energy must be minimized to find the equilibrium configuration 
$\vec R^{(0)}_k$. 
Now consider small displacements $\vec Q_k=\vec R_k-\vec R^{(0)}_k$
around the equilibrium.
The Hamiltonian describing the oscillations of the ions around 
$\vec R^{(0)}_k$ is given by
\begin{equation}
H_{ion}=\frac{1}{2}\sum_{k\mu}M_k\dot Q_{k\mu}^2+ 
\frac{1}{2}\sum_{k\mu,l\nu}\mathcal{H}_{k\mu,l\nu} Q_{k\mu}Q_{l\nu}, 
\nonumber
\end{equation}
where $M_k$ are the are the ionic masses, $\mu,\nu=x,y,z$ denote Cartesian
components of vectors and $\mathcal{H}$ is the Hessian matrix:
$\mathcal{H}_{k\mu,l\nu} = \partial^2 E/ \partial R_{k\mu} \partial R_{l\nu}$.
This can be diagonalized by the transformation
$Q_{k\mu}=\sum_{\alpha=1}^{3N_{vib}} A_{k\mu,\alpha}q_\alpha$,
where $q_\alpha$ are the normal coordinates. Thus, we obtain 
$H_{ion}=\frac{1}{2}\sum_{\alpha}(\dot{q}_\alpha^2+\omega_\alpha^2q_\alpha^2)$,
where $\omega_\alpha$ ($\alpha=1,\ldots,3N_{vib}$) are the 
vibrational frequencies. The transformation matrix $A$ is normalized 
according to 
$A^TMA=1$, $M$ being the mass matrix --- in our case $M$ is simply 
a scalar giving the mass of a gold atom.
Upon using the canonical quantization prescription
$q_\alpha=\left(\hbar/2\omega_\alpha\right)^{1/2}$
$(b_\alpha^\dagger+b_\alpha)$, 
$\dot q_\alpha=\iu\left(\hbar\omega_\alpha/2\right)^{1/2}$
$(b_\alpha^\dagger-b_\alpha)$,
one finally obtains $\op H_{vib}$ in Eq.\ (\ref{e.ham1}).

The electron-vibration interaction may be derived as follows 
\cite{Mahan,ChenZwolakDiVentra03}.
Assume that the electronic single-particle Hamiltonian $H$
is a function of the ionic coordinates, denoted collectively as
$\vec R$. 
Then we may expand
$H_{}(\vec R^{(0)}+\vec Q)\approx $
$H_{}(\vec R^{(0)})+\sum_k\vec Q_k\cdot\vec\nabla_k H_{}|_{\vec Q=0}$.
Defining 
$\op H'_{e}=\sum_{ij}d^\dagger_i\langle i|H_{}(\vec R^{(0)}+\vec Q)|j\rangle d_j$, 
inserting the expansion, and 
using the canonical quantization for $q_\alpha$ again,
one finds $\op H'_{e} = \op H_{e} + \op H_{e-vib}$
[cf. Eq.\ (\ref{e.ham1})], where $H\equiv H(\vec R^{(0)})$, and
the electron-vibration coupling constants 
are given by
\begin{equation}\begin{split} \label{e.ccs}
\lambda_{ij}^\alpha
=\lambda_0\left(\frac{\hbar}{2\omega_\alpha}\right)^{1/2}
\sum_{k\mu}
M_{ij}^{k\mu}A_{k\mu,\alpha},
\end{split}\end{equation}
where
$M_{ij}^{k\mu}=\langle i|\nabla_{k\mu}H_{}|_{\vec Q=0}|j\rangle$.
The calculation of these matrix elements is explained in 
Appendix \ref{s.cccal}.
In Eq.\ (\ref{e.ccs}) we have added a dimensionless factor $\lambda_0$
to describe the strength of the coupling --- in the physical
case $\lambda_0=1$.

\subsection{Propagator formalism}

Use of a local basis allows one to partition the
electronic Hamiltonian and overlap matrices into
parts according to the division in $L$, $C$ and $R$
regions:
\begin{equation}\begin{split}
\gmat{H}=
\left[
\begin{matrix}
\gmat{H}_{LL} & \gmat{H}_{LC} & \gmat{H}_{LR} \\
\gmat{H}_{CL} & \gmat{H}_{CC} & \gmat{H}_{CR} \\
\gmat{H}_{RL} & \gmat{H}_{RC} & \gmat{H}_{RR} \\
\end{matrix}
\right], ~
\gmat{S}=
\left[
\begin{matrix}
\gmat{S}_{LL} & \gmat{S}_{LC} & \gmat{S}_{LR} \\
\gmat{S}_{CL} & \gmat{S}_{CC} & \gmat{S}_{CR} \\
\gmat{S}_{RL} & \gmat{S}_{RC} & \gmat{S}_{RR} \\
\end{matrix}
\right]. \nonumber
\end{split}\end{equation}
Although the dimension of the problem is infinite, 
its single-particle nature allows 
for very effective methods of solution, as long as we may assume that
$\gmat{H}_{RL}=\gmat{H}^T_{LR}\approx0$ and 
$\gmat{S}_{RL}=\gmat{S}^T_{LR}\approx0$, which we shall do.
We shall use the method of non-equilibrium Green functions 
(NEGF). 
In this method, one can restrict the problem only to 
the $C$ part by introducing energy-dependent lead self-energies
which take into account the presence of the semi-infinite
$L$ and $R$ leads in an exact way.

The quantity from which all elastic transport properties
may be extracted, is the retarded Green function of the $C$ part
in the absence of electron-vibration coupling.
We call it $\gmatx{G}^r$, and it may be written as
$\gmatx{G}^r (\epsilon) = [\epsilon\gmat{S}_{CC}-\gmat{H}_{CC}
-\gmat{\Sigma}^r_L -\gmat{\Sigma}^r_R  ]^{-1}$. 
The lead self-energy $\gmat{\Sigma}^r_L$ is given by 
$\gmat{\Sigma}^r_L = \gmat{t}_{CL}\gmat{g}^r_{LL}\gmat{t}_{LC}$, and
$\gmat{\Gamma}_L = \iu (\gmat{\Sigma}^r_L - \gmat{\Sigma}^a_L)$, 
where we define $\gmat{t}_{CL} =\gmat{H}_{CL}-\epsilon\gmat{S}_{CL}$.
The matrix $\gmat{g}^r_{LL}(\epsilon) = [(\epsilon+\iu\gamma_L/2)
\gmat{S}_{LL}-\gmat{H}_{LL}]^{-1}$
is the lead (surface) Green function, where 
$\gamma_L=0^+$.
Similar equations hold for $\gmat{\Sigma}^r_R$.
The lead Green functions $\gmat{g}^r_{LL}$ and $\gmat{g}^r_{RR}$
are ``surface'' Green functions for the semi-infinite leads.
We compute these with the so-called 
decimation technique \cite{GuineaTejedor83}, 
using TB parameters for bulk \cite{BrandbygeKobayashiTsukada99}
in the case of the full $spd$ model. 
The electron-vibration interaction gives rise to further
self-energies, as will be discussed below.

The vibrational modes should in principle be treated in an analogous 
way, by introducing lead self-energies for their propagators. 
However, here we restrict the modes strictly 
to the wire of $N_{ch}$ atoms within the $C$ region 
({i.e.} $N_{vib}=N_{ch}$) and
use the corresponding normal-mode basis for them. 
Thus the number of modes which we have to consider is 
only $3N_{ch}$, and their ``lead coupling'' 
is taken into account only in a phenomenological way.

More details on the propagator technique, including
the expressions for the phonon propagators and 
all self-energy diagrams, are given in Appendix \ref{s.negf}.

\subsection{Calculation of current}
\label{s.current}

The most important physical observable which we are interested in
is the electric (charge) current through the atomic wire,
when a voltage $V$ is applied. 
We denote $eV=\mu_L-\mu_R$,
where $\mu_{L,R}$ are the $L$ and $R$ side chemical potentials,
and $e>0$ is the absolute value of electron charge.
We also define $f_{L,R}(\epsilon)=f(\epsilon-\mu_{L,R})$, 
where $f(\epsilon)=1/[\exp(\beta\epsilon)+1]$ is the Fermi function, 
and $\beta=1/k_BT$ is the inverse temperature.

It may be shown that the current flowing 
through the interface from $L$ to $C$ ($C$ to $R$)
in stationary state is given by (see Appendix \ref{s.cder})
\begin{equation} \label{e.curr0}
I^\Omega=\pm\frac{2e}{\hbar}
\int\frac{\upd\epsilon}{2\pi}
\Tr[\gmat{G}_{C\Omega}^{+-}(\epsilon)\gmat{t}_{\Omega C}(\epsilon)
-\gmat{t}_{C\Omega}(\epsilon)\gmat{G}_{\Omega C}^{+-}(\epsilon)],
\end{equation}
where $\Omega=L(R)$ is chosen with the upper (lower) sign,
and the factor $2$ accounts for spin degeneracy.
The Green functions $\gmat{G}^{+-}$ are 
defined in Appendix \ref{s.negf} by Eq.\ (\ref{e.gdef}). 
Developing Eq.\ (\ref{e.curr0}) further, 
it is convenient to split it into 
two parts: $I^{L,R}=I_{el}+I_{inel}^{L,R}$, where
\begin{equation}\begin{split} \label{e.curr}
I_{el} = \frac{2e}{\hbar}\int\frac{\upd\epsilon}{2\pi} &
\Tr[\gmat{G}^r\gmat{\Gamma}_R\gmat{G}^a\gmat{\Gamma}_L](f_L-f_R) \\
I_{inel}^{L,R}  = \pm\frac{2e}{\hbar}\iu 
\int\frac{\upd\epsilon}{2\pi}& \Tr\{\gmat{G}^a \gmat{\Gamma}_{L,R}\gmat{G}^r\\
&\times[(f_{L,R}-1)\gmat{\Sigma}^{+-}_{e-vib}-f_{L,R}\gmat{\Sigma}^{-+}_{e-vib}]\}.
\end{split}\end{equation}
Here we define the full retarded and
advanced Green functions $\gmat{G}^{r,a}$, where
$\gmat{G}^r = [\epsilon\gmat{S}_{CC}-\gmat{H}_{CC}$
$-\gmat{\Sigma}^r_L -\gmat{\Sigma}^r_R -\gmat{\Sigma}^r_{e-vib} ]^{-1}$
and $\gmat{G}^a=[\gmat{G}^r]^\dagger$.
The new self-energies $\gmat{\Sigma}^r_{e-vib}$ and 
$\gmat{\Sigma}^{\pm\mp}_{e-vib}$
are due to the electron-vibration interaction, and
they are discussed in more detail in Appendix \ref{s.negf}.
Since they vanish in the absence of $\gmat{\lambda}^\alpha$,
we call the $I_{inel}^{L,R}$ part an ``inelastic'' current, 
while $I_{el}$ is the ``elastic'' part \cite{GalperinRatnerNitzan04}.

If we do lowest-order perturbation theory with respect to $\gmat{\lambda}^\alpha$,
we may expand $\gmat{G}^r =\gmatx{G}^r + \gmatx{G}^r \gmat{\Sigma}^r_{e-vib} \gmatx{G}^r +\cdots$.
In this way the elastic current is split into two parts
as $I_{el}=I_{el}^0+\delta I_{el}$, where $\delta I_{el}$
is an ``elastic correction''. We find
\begin{equation}\begin{split} \label{e.pertcurr}
I^0_{el} = \frac{2e}{\hbar}\int\frac{\upd\epsilon}{2\pi}&
\Tr[\gmatx{G}^r\gmat{\Gamma}_R\gmatx{G}^a\gmat{\Gamma}_L](f_L-f_R) \\
\delta I_{el}  = \frac{4e}{\hbar}\int\frac{\upd\epsilon}{2\pi}
&\re\Tr[\gmat{\Gamma}_L \gmatx{G}^r\gmat{\Sigma}^r_{e-vib}\gmatx{G}^r
\gmat{\Gamma}_R\gmatx{G}^a] \\
&\times(f_L-f_R) \\
I_{inel}^{L,R}  = \pm\frac{2e}{\hbar}\iu
\int\frac{\upd\epsilon}{2\pi}& \Tr\{\gmatx{G}^a \gmat{\Gamma}_{L,R}
\gmatx{G}^r \\
&\times[(f_{L,R}-1)\gmat{\Sigma}^{+-}_{e-vib}
-f_{L,R}\gmat{\Sigma}^{-+}_{e-vib}] \}.
\end{split}\end{equation}
A proof of current conservation, 
that is $I_{}^L=I_{}^R\equiv I$,
is sketched in Appendix \ref{s.conser}.

Besides the charge current, other interesting observables would be the 
heat current (or power dissipation)
\cite{Frederiksen04a}, current
noise \cite{ChenDiVentra05}, and possibly spin current in case 
of magnetic materials. 
We only consider the charge current here, as it is
the only one which can easily be measured.
More specifically, we shall be interested in
the differential conductance $G(V)=\upd I/\upd V$
and its derivative, since these quantities reveal the
signatures of the vibrational-mode coupling most clearly.

\section{Wide-band limit} \label{s.wbl}

Even in the case of the perturbative current formulas [Eqs.\ (\ref{e.pertcurr})],
the expressions will involve double energy integrals which can be
very cumbersome to evaluate. These general formulas are discussed more in
Appendix \ref{s.form}, where they are rewritten in terms of 
distribution functions and energy-dependent 
transport coefficients. 
However, the existence of different energy scales
in the problem allows us to make an important simplification.

The energies of the vibrational modes are on the order of $10$ meV,
so that we are only interested in the differential conductance
for voltages up to $V\approx40$ mV, at most.
Together with the temperature $T\approx4.2$ K, this
determines the width of the energy window
around the Fermi energy ($\epsilon_F$) which is important for transport.
However, for the atomic wires which we are considering,
the electronic density of states tends to vary at the much larger 
energy scales $\sim 1$ eV around $\epsilon_F$. Thus, to a good approximation, 
we may neglect this energy dependence, and simply evaluate all 
the electronic Green functions at the Fermi energy.
This approximation is often called the ``wide-band limit'' (WBL).

\subsection{Current}

In the WBL approximation, the current expressions
of Appendix \ref{s.form} may 
simplified considerably, since some of the energy integrals 
may be done analytically.
In this case the current $I$ is easily divided
into \emph{symmetric} and \emph{asymmetric}
parts, according to the symmetry 
of their contributions to $G(V)$
under the reversal of $V$ \cite{Paulsson05}.
Thus $I=I^{sym}+I^{asy}$, where the symmetric current
is
\begin{equation}\begin{split} \label{e.isym}
I^{sym}&=\frac{2e^2}{h}T_0 V +
\frac{2e}{h}
\sum_\alpha
\int_0^{\infty}\upd\omega_1\rho_\alpha(\omega_1)
\bigg\{ \\
&[(T_\alpha^{ec}+T_\alpha^{in})(2N_\alpha(\omega_1)+1)
-(T_\alpha^{ecLR}+T_\alpha^{in})]eV \\
+&(T_\alpha^{ecLR}+T_\alpha^{in})
\bigg(
\frac{\omega_1-eV}{e^{\beta(\omega_1-eV)}-1}-
\frac{\omega_1+eV}{e^{\beta(\omega_1+eV)}-1}
\bigg)
\bigg\} 
\end{split}\end{equation}
and the asymmetric part is
\begin{equation}\begin{split}
I^{asy}&=-\frac{2e}{h}\sum_\alpha\int_{-\infty}^{\infty}\upd\omega_1
\frac{1}{2\pi}\re[d^r_{\alpha}(\omega_1)]
T^{asy}_\alpha \\
\times&\bigg[
2n(\omega_1)\omega_1 
- \frac{\omega_1-eV}{e^{\beta(\omega_1-eV)}-1} 
- \frac{\omega_1+eV}{e^{\beta(\omega_1+eV)}-1} 
\bigg].
\end{split}\end{equation}
Here $n(\epsilon)=1/[\exp(\beta\epsilon)-1]$ is the 
Bose distribution and
$N_\alpha$ is the voltage-dependent 
mode distribution, to be discussed shortly.
The function $\rho_\alpha$ is the 
vibrational density of states (DOS), given in general by
Eq.\ (\ref{e.fullphdos}).
We approximate it here by using the imaginary part of $d^r_\alpha$
in Eq.\ (\ref{e.bared}), that is
\begin{equation}\begin{split} \label{e.apprphdos}
\rho_\alpha(\epsilon)=
\frac{1}{\pi}\frac{\eta/2}{(\epsilon-\hbar\omega_\alpha)^2+\eta^2/4}
-\frac{1}{\pi}\frac{\eta/2}{(\epsilon+\hbar\omega_\alpha)^2+\eta^2/4},
\end{split}\end{equation}
where we take $\eta$ as a finite phenomenological
parameter describing the effect of coupling the 
vibrational modes to an external bath. This bath is
provided by the leads \cite{GalperinRatnerNitzan04}.
However, we are neglecting any renormalizations of the
bare frequencies $\omega_\alpha$, so that the
main purpose of $\eta$ here is to broaden the DOS. 
Similarly, $\re d^r_\alpha$ is obtained from 
Eq.\ (\ref{e.bared}). 

The current involves a number of transport
coefficients, which determine the shape of 
the current-voltage characteristics.
The coefficients of $I^{sym}$
may be computed as follows:
\begin{equation}\begin{split} \label{e.t0x}
T_0
&= \Tr[\gmatx{G}^r\gmat{\Gamma}_R\gmatx{G}^a
\gmat{\Gamma}_L]|_{\epsilon_F}, \\
\end{split}\end{equation}
\begin{equation}\begin{split} \label{e.tinx}
T_{\alpha}^{in}
&= \Tr[\gmatx{G}^r
\gmat{\Gamma}_R\gmatx{G}^a\gmat{\lambda}^\alpha
\gmatx{G}^a\gmat{\Gamma}_L\gmatx{G}^r\gmat{\lambda}^\alpha]|_{\epsilon_F},\\
\end{split}\end{equation}
\begin{equation}\begin{split} \label{e.tecx}
T_{\alpha}^{ec}
&=2\re\Tr[
\gmatx{G}^r\gmat{\Gamma}_R\gmatx{G}^a
\gmat{\Gamma}_L\gmatx{G}^r\gmat{\lambda}^\alpha
\gmatx{G}^r\gmat{\lambda}^\alpha]|_{\epsilon_F}, \\
\end{split}\end{equation}
\begin{equation}\begin{split} \label{e.teclrx}
T_{\alpha}^{ecLR}&=T_\alpha^{ecL}+T_\alpha^{ecR}\\ 
&=\re\Tr[\gmatx{G}^r\gmat{\Gamma}_R\gmatx{G}^a
\gmat{\Gamma}_L\gmatx{G}^r\gmat{\lambda}^\alpha
(\gmatx{G}^r-\gmatx{G}^a)
\gmat{\lambda}^\alpha]|_{\epsilon_F}. \\
\end{split}\end{equation}
Here $T^{ecLR}_\alpha$ 
is a sum of the two coefficients 
$T^{ecL}_\alpha$ and $T^{ecR}_\alpha$ defined
in Eqs.\ (\ref{e.eccoeffs}).
The coefficient $T^{in}_\alpha$ is always positive, 
while $T^{ecLR}_\alpha$ and $T^{ec}_\alpha$ can apparently have either sign.
The latter two represent interferences between various processes, 
and are responsible for the enhanced backscattering
needed for the conductance drops.
The asymmetric coefficient
\begin{equation}\begin{split}
T^{asy}_\alpha=\re\Tr[\gmatx{G}^r
\gmat{\Gamma}_R\gmatx{G}^a\gmat{\Gamma}_L\gmatx{G}^r
\gmat{\lambda}^\alpha
\gmatx{G}^r(\gmat{\Gamma}_R-\gmat{\Gamma}_L)\gmat{G}^a
\gmat{\lambda}^\alpha]|_{\epsilon_F}
\end{split}\end{equation}
vanishes for sufficiently symmetric junctions, making $I^{asy}=0$.
For us, this is always the case, even 
with zigzag chains.

\subsection{Mode distribution}

Our approximation is essentially that of lowest-order perturbation
theory in the coupling constant $\gmat{\lambda}^\alpha$. 
However, as explained in Appendix \ref{s.negf}, there is
a natural way of extending the theory somewhat further
by taking into account the ``phonon polarizations'' 
in a voltage-dependent distribution function $N_\alpha$.
This is given by Eq.\ (\ref{e.distrib}), i.e.
\begin{equation}\begin{split} \label{e.modedist}
N_\alpha(\epsilon)
&=-\frac{1}{2}\frac{\im\Pi^{+-}_\alpha(\epsilon)+
n(\epsilon)\eta\epsilon/\hbar\omega_\alpha}
{\im\Pi^r_\alpha(\epsilon)-\eta\epsilon/2\hbar\omega_\alpha}, \\
\end{split}\end{equation}
where 
$\eta$ is the same bath-coupling parameter 
as in $\rho_{\alpha}$.
Here
$\im\Pi^{+-}_\alpha$ and
$\im\Pi^r_\alpha$ 
are imaginary parts of the phonon polarizations
(see Appendix \ref{s.negf}).
In the WBL one may show that
\begin{equation}\begin{split} \label{e.wblpipm}
2\pi \im&\Pi^{+-}_\alpha(\epsilon) \\
\approx &
\Tr[\gmat{\lambda}^\alpha\gmatx{G}^r\gmat{\Gamma}_L
\gmatx{G}^a\gmat{\lambda}^\alpha\gmatx{G}^r\gmat{\Gamma}_R 
\gmatx{G}^a]|_{\epsilon_F} \\
&\times\bigg[
\frac{\epsilon+eV}{e^{\beta(\epsilon+eV)}-1}
+
\frac{\epsilon-eV}{e^{\beta(\epsilon-eV)}-1}
\bigg] \\
+&
\bigg\{
\Tr[\gmat{\lambda}^\alpha\gmatx{G}^r\gmat{\Gamma}_L
\gmatx{G}^a\gmat{\lambda}^\alpha\gmatx{G}^r\gmat{\Gamma}_L
\gmatx{G}^a]|_{\epsilon_F} \\
&+
\Tr[\gmat{\lambda}^\alpha\gmatx{G}^r\gmat{\Gamma}_R
\gmatx{G}^a\gmat{\lambda}^\alpha\gmatx{G}^r\gmat{\Gamma}_R 
\gmatx{G}^a]|_{\epsilon_F} 
\bigg\} 
\frac{\epsilon}{e^{\beta\epsilon}-1}
\end{split}\end{equation}
and
\begin{equation}\begin{split} \label{e.wblpir}
\im \Pi^{r}_\alpha(\epsilon)\approx
-(\epsilon/\pi)\Tr[\gmat{\lambda^\alpha}\im\gmatx{G}^r
\gmat{\lambda^\alpha}\im\gmatx{G}^r],
\end{split}\end{equation}
which is proportional to the electron-hole damping rate of Ref.\ 
\onlinecite{Paulsson05}.
Using
$\gmatx{G}^r(\gmat{\Gamma}_L+\gmat{\Gamma}_R)\gmatx{G}^a$
$=-2\im\gmatx{G}^r$,
one may easily show that for $V=0$ 
Eqs.\ (\ref{e.modedist})-(\ref{e.wblpir}) 
indeed yield $N_\alpha(\epsilon)=n(\epsilon)$
for any $\eta$. 

Here it is important to note a few things. 
In the expression 
for the distribution function, the limit $\eta\rightarrow 0+$
corresponds to the case where the vibrational modes 
are uncoupled from leads.
Supposing that one also wishes to take
the phonon polarizations to zero, which is formally accomplished by
taking $\lambda_0\rightarrow 0$,
one discovers an interesting thing: the two limits do not commute.

If we take first the limit $\eta\rightarrow 0$, 
then the result actually becomes
independent of $\lambda_0$, 
since 
$\Pi^r_\alpha,\Pi^{+-}_\alpha\propto\lambda_0^2$ and 
the $\lambda_0^2$-factors cancel.
A physical interpretation can be described as follows.
If the vibrational modes are not coupled to
any external bath, then even an infinitesimally small
coupling constant can eventually lead to a \emph{stationary state}
with a strongly nonequilibrium mode occupation. Here
emission and absorption of phonons are in balance, 
and hence there is no net energy transfer between the
electrons and the vibrations. 
Following Ref.\ \onlinecite{Frederiksen04a}, 
we call this the \emph{externally undamped limit}, 
although our way of computing $N_\alpha$ is quite different.
In this case the voltage-dependence of $N_\alpha(\hbar\omega_\alpha)$
shows a sharp kink at $V=\hbar\omega_\alpha/e$, and a subsequent
linear increase \cite{Frederiksen04a}.

In the opposite case, where
$\lambda_0\rightarrow 0$ first, the expression becomes
independent of $\eta$, and we recover the Bose distribution. 
This corresponds to the limit where the vibrational modes are strongly
damped by coupling to a heat bath which is in equilibrium. 
This is the \emph{externally damped limit}.
However, for a finite $\gmat{\lambda}^\alpha$ this limit can only be 
reached with a large enough finite $\eta$. Thus the externally damped 
limit should also imply a considerable
broadening of the vibrational modes.

Note that in both of the above limits, $N_\alpha$ is zeroth
order in $\lambda_0$.
In these two cases our expression for $I^{sym}$ [Eq.\ (\ref{e.isym})] 
is indeed of second order in $\lambda_0$, and there
should be no corrections within that order.
In general, however, Eq.\ (\ref{e.modedist}) 
for $N_\alpha$ generates terms of all orders in $\lambda_0^2$.
These are not, strictly speaking, warranted, because they
represent only a small subset of all possible 
higher-order terms in the current.

\subsection{Further discussion}

Apart from the $\rho_\alpha$-weighted integral,
the only difference between Eq.\ (\ref{e.isym}) and the approximation of 
Ref.\ \onlinecite{Paulsson05} for $I^{sym}$
is that $T_\alpha^{ec}\neq T_\alpha^{ecLR}$.
The difference $T_\alpha^{ec}-T_\alpha^{ecLR}$ is 
proportional to $\re \gmatx{G}^r$, and is typically very 
small. It could well be neglected.
If $N_\alpha\equiv$constant, then
the direction and size of the conductance step 
is solely determined by the combination 
$T^{ecLR}_\alpha+T^{in}_\alpha$.
Since $T_\alpha^{in}$ is positive,
the $I_{inel}$ part of the current 
always tends to \emph{increase}
the conductance.
This can be seen as a result of the appearance
a new, inelastic conduction ``channel'' when $eV>\hbar\omega_\alpha$.
However, numerically one finds that the 
coefficient
$T^{ecLR}_\alpha$, due to interferences
between various elastic processes, 
is typically negative and 
$|T^{ecLR}_\alpha|>T_\alpha^{in}$ when $T_0\approx 1$. 
Thus, the net effect of
the vibrational coupling is a \emph{decrease}
in the conductance.
In general, the voltage-dependence of $N_\alpha$ also 
affects the shape of $G(V)$.
Since 
$T^{ec}_\alpha+T^{in}_\alpha$ is usually also 
negative, the effect of local heating is 
to give a finite negative slope to $G(V)$ after
the step. There can also be an increase
in the apparent height of the step. 

By using the vibrational DOS of Eq.\ (\ref{e.apprphdos}),
we are neglecting broadenings and frequency shifts due 
to the electron-vibration coupling.
The shifts are given by the quantities $\re \Pi^r_\alpha$ 
and, using Eq.\ (\ref{e.phself}), we estimate them to be on the 
order of $-1$ meV.
Neglecting the broadenings is certainly justified, since 
they are typically on the order of
$|\!\im\Pi^r_\alpha(\hbar\omega_\alpha)|/\hbar\omega_\alpha\lesssim 10^{-2}$,
which is much smaller than the usual $k_BT/\hbar\omega_\alpha$
due to temperatures $T\sim1$ K \cite{Frederiksen04a}. 
If we assume also $\eta$
to be small compared to the other relevant energy scales,
i.e. if 
$|\!\im\Pi_\alpha^r|,\eta\ll k_BT\ll\hbar\omega_\alpha$,
then
$\rho_\alpha(\omega_1)\approx\delta(\omega_1-\hbar\omega_\alpha)-\delta(\omega_1+\hbar\omega_\alpha)$,
and the $\omega_1$ integral 
in Eq.\ (\ref{e.isym}) may be done analytically. 
The calculation of the lead-coupling effects is difficult, 
but it is very possible that they can lead to
broadenings $\eta/\hbar\omega_\alpha\sim1$.
Thus, the validity of the delta-function approximation of 
$\rho_\alpha$ may be questioned.
Naturally, the lead coupling can also shift the
vibrational frequencies.

Note also that 
$N_{\alpha}(\epsilon)$ diverges
if $\beta\epsilon\rightarrow 0$.
Since in Eq.\ (\ref{e.isym}) $N_{\alpha}(\epsilon)$
is evaluated at $\epsilon\sim\hbar\omega_\alpha$,
there can be an arbitrarily large renormalization 
of the zero-bias conductance
if there are very low-frequency vibrational modes
($\hbar\omega_\alpha\sim k_BT$)
with a strong coupling to the electrons.
In such a case, the theory appears to  
break down.
With linear wires the condition 
$k_BT\ll\hbar\omega_\alpha$ 
is easily satisfied
for all strongly coupled modes $\alpha$, 
but in case of the zigzag-chains (and/or other materials) 
it may not be.
In any case, we limit our study mostly to the linear wires 
in this paper. 
We note that these restrictions are present also in
Ref.\ \cite{Paulsson05}, where $\eta=0^+$.
In fact, in our formulation the problem is perhaps partly corrected 
by the presence of the ``broadening factor'' $\rho_\alpha$.
In general $\int_0^\infty\upd\omega_1\rho_\alpha(\omega_1)<1$, 
and when $\hbar\omega_\alpha\rightarrow 0$, the whole function 
tends to zero. Thus low-frequency modes contribute to the
current with a very small weight.

\section{Simple chain models} \label{s.simple}

Here we shall first present some example results, using 
an adaptation of the simple 
chain models of Refs.\ \onlinecite{SuSchriefferHeeger79} and 
\onlinecite{Frederiksen04b}.
Only after this we turn to the full $spd$-tight binding parametrization.
\begin{figure}[!tb] 
\includegraphics[width=0.65\linewidth]{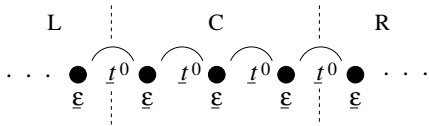}
\caption{Infinite multi-orbital nearest-neighbor chain in 
equilibrium, where the sites are separated by a distance $a$.} 
\label{f.hopchain}
\end{figure}
Chain models of this kind are very appealing, 
because they can be studied
analytically to a large extent,
and allow us to make our main points in a simple fashion.

The single-particle Hamiltonians $\gmat{H}$ which we consider, are
of the block-tridiagonal form
\begin{equation}\begin{split}
\gmat{H}=\left[\begin{matrix}
\ddots &\ddots & \ddots & & \\
& \ul{t}_{i,i-1} & \ul{\epsilon} & \ul{t}_{i,i+1} & \\
& & \ddots &\ddots & \ddots \\
\end{matrix}\right],
\end{split}\end{equation}
graphically depicted in Fig.\ \ref{f.hopchain}.
Here $\ul{\epsilon}$ includes the on-site energies, and
$\ul t_{i,i+1}$ are the inter-site hopping matrices
between sites $i$ and $i+1$. They 
are modulated as a function
of the longitudinal atomic displacements $Q_i$ as
$\ul{t}_{i,i+1}(Q)=\ul{t}^0+\ul{t}'(Q_{i}-Q_{i+1})$. 
We only consider orthogonal tight-binding here, such that $\gmat S=\gmat 1$. 
The chain is split into three parts, where the 
$C$ part has $N_{ch}$ atoms. The $L$ and $R$ ``leads'' 
are semi-infinite chains. The displacements $Q_i$
are restricted to the $C$ part, where 
the Hessian is assumed to be of the form
\begin{equation}\begin{split}
\mathcal{H}=K\left[\begin{matrix}
\ddots &\ddots & \ddots & & \\
& -1 & 2 & -1 & \\
& & \ddots &\ddots & \ddots \\
\end{matrix}\right],
\end{split}\end{equation}
with fixed boundary conditions at the ends \cite{Frederiksen04b}.
Unlike in the $spd$ model to be explained below,
here the spring constant $K$ is taken 
as a new, separate parameter.
The vibrational modes are simply obtained by diagonalizing this 
$\mathcal{H}$.

\subsection{Single ($s$) orbital: discussion}

To remind ourselves of some of the well-known arguments, 
let us first discuss the simple single-orbital
orthogonal model of Ref.\ \cite{Frederiksen04b}. 

If we assume that charge neutrality is achieved by one 
electron per atomic site, then the Fermi energy $\epsilon_F$ lies exactly
in the middle of the band $-2|t_0|<\epsilon-\epsilon_0<2|t_0|$, 
where $\epsilon_0$ is the on-site energy, and
$t_{0}$ is the inter-site hopping.
The Fermi wavevector is then $k_F=\pi/2a$, where
$a$ is the inter-atomic distance.
Since the chain has 
full translational invariance, momentum conservation dictates that
only vibrational modes with the wave vector 
$q_{vib}=2k_F=\pi/a$ may 
be excited, via the backscattering of Fermi-point electrons.
Furthermore, if the atomic orbitals are invariant with respect to rotations
around the axis of the chain ($s$-wave, say) the modes can only be longitudinal
ones. As it happens, $q_{vib}=\pi/a$ corresponds exactly to
the Brillouin zone boundary and thus to the highest-frequency modes.

More technically, the momentum conservation 
can be seen to follow from the form of the
Fourier-transforms $M_{k_1,k_2}^q\propto\delta_{k_1-k_2,q}$
of the coupling matrix elements $M_{ij}^k$ \cite{Mahan}.
We note that the momentum conservation can only be strictly valid if
one considers the actual vibrational modes of the whole infinite 
chain. Interestingly, it appears to 
remain approximately valid even if the vibrational modes 
are restricted only to a small, finite part of the chain
\cite{Frederiksen04b}.
Indeed, as found in 
Ref.\ \onlinecite{Frederiksen04b}, 
when the Fermi energy is in the
middle of the band, there is only a single visible step in the conductance.
But, as we shall discuss next, this appears to be somewhat accidental.

\subsection{Two ($s$ and $p_z$) orbitals}

\begin{figure}[!tb] 
\includegraphics[width=0.9\linewidth,clip=]{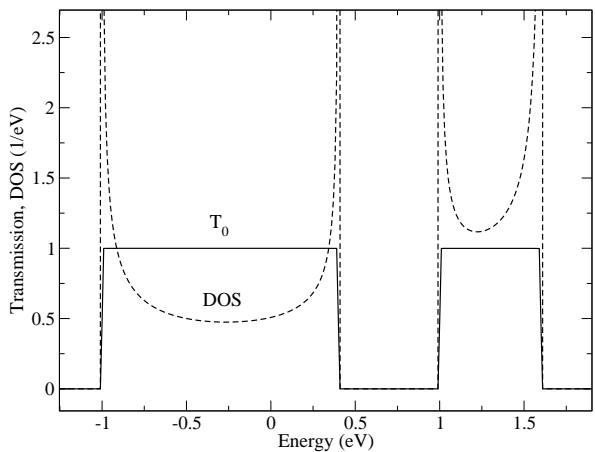}
\caption{Elastic transmission $T_0$ (solid line) 
and density of states (dashed line)
for an $sp_z$ chain 
model. 
} 
\label{f.SPZTRANS}
\end{figure}

\begin{figure}[!tb] 
\includegraphics[width=0.99\linewidth,clip=]{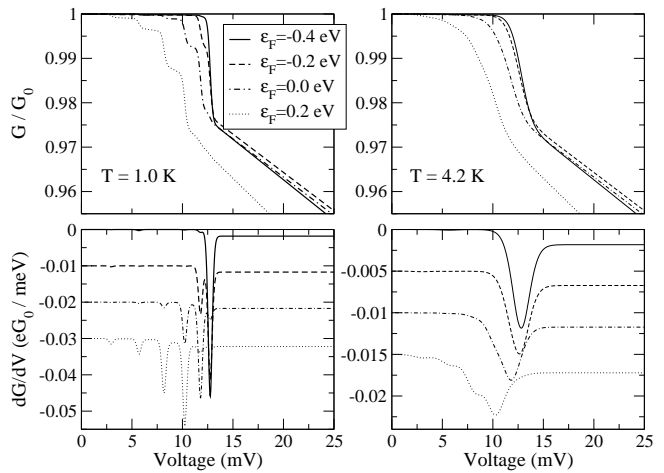}
\caption{The top panels show the conductance vs. voltage
$G(V)$ for the $sp_z$ chain model with parameters as in 
Table \ref{t.spzpar}, and the lower panels show the 
derivative $\upd G/\upd V$.
Here the curves have been shifted by 
integer multiples of 0.01 or 0.005 for clarity.
The left panels are at $T=1.0$ K, while the right
panels are at $T=4.2$ K.
The vibrational modes are as in Fig.\ 1 of 
Ref.\ \onlinecite{Frederiksen04b}.} 
\label{f.SPZCHAIN}
\end{figure}

Although such a single-orbital model can correctly describe
the most important experimental observations, it neglects 
some details of realistic gold wires. 
Although atomic gold contacts and chains
typically have only one fully open conduction channel at the Fermi surface,
this channel is actually formed from the hybridization on multiple orbitals
with rotational symmetry around the chain axis 
($s$, $p_z$, $d_{3z^2-r^2}$).
Thus we consider here a simple generalization of the 
above chain model to the case of \emph{two} orbitals, 
with $s$ and $p_z$ characters, respectively. 
In Table \ref{t.spzpar} we show a set of example parameters
for this model.
\begin{table}[!htb]
\begin{tabular}{lll}
\hline
Quantity & Symbol & Value \\
\hline
\hline
Number of chain atoms    & $N_{ch}$        & 6 \\
$s$-orbital energy       & $\epsilon_{ss}$ & 0.0 eV  \\
$p$-orbital energy       & $\epsilon_{pp}$ & 1.0 eV  \\
Bare $s-s$ hopping       & $t_{ss}^0$ & -0.5 eV \\
Bare $p-p$ hopping       & $t_{pp}^0$ & 0.3 eV \\
Bare $s-p$ hopping       & $t_{sp}^0=-t_{ps}^0$ & 0.35 eV \\
$s-s$ hopping modulation & $t'_{ss}$  & -0.3 eV \\
$p-p$ hopping modulation & $t'_{pp}$  & 0.3 eV \\
$s-p$ hopping modulation & $t'_{sp}=-t'_{ps}$  & 0.3 eV \\
Fermi energy       & $\epsilon_F$ & -0.4---0.2 eV \\
Atomic Mass        & $M$   & 197 a.m.u. \\
Spring constant    & $K$   & 2.0 eV/\AA$^2$ \\
Temperature        & $T$   & 1.0---4.2 K \\
Phonon broadening  & $\eta$ & 0.002 meV \\
\hline
\end{tabular}
\caption{Parameters for the $sp_z$ schain.
Here $t^0_{ij}$ are elements of the matrix 
$\ul{t}^0$, for example.} \label{t.spzpar}
\end{table}
The resulting transmission and DOS curves are shown in 
Fig.\ \ref{f.SPZTRANS}
and the conductance-voltage characteristics 
in Fig.\ \ref{f.SPZCHAIN}.
The density of states is defined as
$\mathcal{D}_i(\epsilon)=-(1/\pi)\sum_\alpha\im G_{i\alpha,i\alpha}(\epsilon)$,
where $i$ now stands for atomic sites and $\alpha$ for orbitals.

The results are very similar to those of the 
single-orbital chain \cite{Frederiksen04b}, but the situation is a bit closer to 
what happens in the full $spd$ model.
For example, the hybridization of $s$ and $p_z$ result in a 
\emph{gap} in the DOS. If we still populate
the chain with one electron per site, the Fermi energy 
will again be in the middle of the lower band (the ``$s$ band''),
and there is only a single drop in the conductance. This
is illustrated by the $\epsilon_F=-0.4$ eV case in 
Fig.\ \ref{f.SPZCHAIN}. 
But for $\epsilon_F$'s deviating from the center of the band, 
we find that the conductance step generally consists of several substeps, 
although the total step height remains almost the same.
Notice that if we were to use two electrons per site, 
then the Fermi energy would lie in the gap, and the transmission would be 
zero. It must be noted, however, that the assumption of the
WBL will break down if the Fermi energy is very close to a band edge.

Within this model of a gold chain, it is probably
most physical to occupy the atoms with a single electron per site. 
However, in a more realistic description the $s$ and $p_z$
orbitals hybridize also with lower-energy $d$ orbitals. 
Also, the real 
systems are not translationally invariant, and 
charge neutrality need not be fully satisfied locally. 
Such things can complicate the picture enormously. 
Furthermore, the
arbitrariness of the charge-neutrality procedure in the
full $spd$ model below renders the position of the 
Fermi energy with respect to the electronic structure
of the wire somewhat uncertain.
Thus, as we shall see, all of the cases shown in 
Fig.\ \ref{f.SPZCHAIN} are actually good descriptions of our 
results for the full $spd$ model.

Although at low temperature there can be several peaks, at higher 
temperatures the steps fuse together, since the vibrational
frequencies of the different modes are quite close to each other. 
The fusing effect of the steps would be further enhanced, if we introduced a 
larger broadening $\eta$ for the vibrational modes -- here 
$\eta=0.002$ meV $\ll k_B T$. The small $\eta$ also
affects the results in another way: since the ``lead coupling'' of vibrations
is small, the wire heats, and the distribution functions $N_\alpha$
differ strongly from the Bose distribution. The signature of this
heating is the steep downward slope of the $G(V)$ curves 
at high voltage \cite{Frederiksen04a}.

We have also tested the effect of $n$'th-nearest-neighbor hoppings
and addition of nonorthogonality, but these have no essential qualitative 
effect on the results. These will be taken into account
in the full $spd$ parametrization, which we now turn to.

\section{Full $spd$ parametrization} \label{s.spd}

In this section we describe a more realistic tight-binding
approach to the problem. We use the nine-orbital $spd$
parametrization of Papaconstantopoulos 
et al.\ \cite{CohenMehlPapa94,MehlPapa96,PapaMehl03,website}.
This type of $spd$ TB approach is 
known to reproduce very well some nontrivial \emph{ab initio} results, like 
the numbers of conduction channels and
the formation of zigzag Au chains \cite{SanchezPortal99,daSilvaNovaes04}.
Thus we can be confident that the method gives at least
good order-of-magnitude estimates for all of the quantities 
which we shall be interested in.

However, since the parameters are extracted from first-principles \emph{bulk}
calculations, they cannot be exactly correct for atomic point contacts,
where the important atoms of the structure are significantly less 
coordinated than in bulk. 
It has thus become customary in the method to ``correct'' 
the parameters in the 
central cluster in order to satisfy local charge neutrality. 
Doing this typically brings the 
central cluster levels better in resonance with the lead orbitals.
We compute the charge with the so-called Mulliken population 
analysis (see Appendix \ref{s.cder}), 
and only shift the on-site energies of the Hamiltonian.
Tests with other ways of achieving neutrality give very similar 
results \cite{BrandbygeKobayashiTsukada99}.
Furthermore, we have compared to results obtained without charge 
neutrality, and again find that there is not much qualitative difference, 
although charge neutrality yields conductances closer to $G_0$ on 
average.
Also, without charge neutrality there appears to be a tendency for 
seeing lower-frequency vibrational modes in the conductance 
curves.

The results of this paper are generated using finite 
lead broadenings $\gamma_{L,R}\approx1.0$ eV. 
The limit $\gamma_{L,R}\rightarrow 0^+$
can in principle be taken without affecting the results in any 
essential way.

\subsection{Geometry optimization and vibrational modes}

\begin{figure}[!tb] 
\includegraphics[width=0.5\linewidth,clip=]{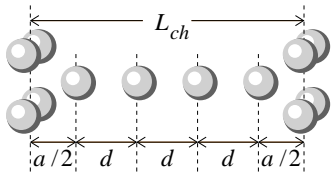}
\caption{Dimensions of the unoptimized geometry 
with $N_{ch}=4$. Here $a=4.08$ \AA, the fcc lattice constant.
Only the coordinates of the $N_{ch}$ chain atoms are optimized.} 
\label{f.chaindims}
\end{figure}

We consider two types of ideal geometries, the ``A'' and ``B'' ones,
shown in Figs.\ \ref{f.lcr} and \ref{f.lcr_pyra}, respectively. 
As mentioned, the
leads are assumed to be of fcc type with the [001]
axis in the transport (or $z$) direction.
Before geometry optimization, 
the chain atoms are positioned as described in 
Fig.\ \ref{f.chaindims}.
The ``length of the wire'' $L_{ch}$ is defined
as $L_{ch}=a+d(N_{ch}-1)$, where $N_{ch}$ is the number of atoms in the
chain, $d$ the distance between them, and $a=4.08$ \AA $\;$ is 
the equilibrium lattice constant of the bulk fcc lattice.
We only optimize the geometry of the $N_{ch}$ chain atoms --- also 
in geometry B which has the ``pyramids''.
Thus, although the interatomic
distances change slightly from those of Fig.\ \ref{f.chaindims},
$L_{ch}$ remains fixed.

To estimate the total energy $E(\vec R_k)$, 
we simply take a cluster which includes the wire and
some atoms from the leads, solve for the electronic eigenstates 
$\epsilon_\alpha$, and then occupy the states according to charge neutrality.
This energy, as a function of the $3N_{ch}$ wire coordinates, 
is then optimized with standard library routines.
As known previously \cite{SanchezPortal99}, it is 
often energetically favorable for the gold 
chains to exist in a zigzag-like pattern instead of a linear one.
Only after a sufficient amount of stretching (i.e., with a larger $d$) 
does the linear configuration become stable, after which it
remains linear until the wire breaks.
We find that the maximum $d$ at breaking is, 
depending on the geometry, typically something between 2.70--2.85 \AA.
There is no clear trend with increasing $N_{ch}$.

After the geometry is optimized, the energy function 
is used to compute the Hessian matrix $\mathcal{H}$. 
The eigenvalues 
$k_\alpha$ ($\alpha=1,\ldots,3N_{ch}$) 
are all positive, and the vibrational frequencies are simply given by
$\omega_\alpha=\sqrt{k_\alpha/M}$.
With both geometries, A and B, we obtain quite similar 
vibrational frequencies and modes. For a linear wire,
the modes can be classified as longitudinal or transverse 
in character. The highest-frequency modes are then always 
longitudinal ones, and the highest of them is of the
ABL type \cite{Frederiksen04a}.

\subsection{Elastic transmission}

Perhaps the most characteristic experimental property
of gold chains is that they appear to have a conductance
very close to the quantum of conductance $G_0$.
Thus we shall briefly comment on the elastic transmission
properties of the chains in our calculations.
The present TB method was previously only used with bulk distances
$d\approx2.885$ {\AA} 
between all the atoms \cite{BrandbygeKobayashiTsukada99}.
In this case, we find very similar transmission functions
$T_0(\epsilon)$ for our geometry B.
These are often characterized by 
very long plateaus (of widths up to $2-3$ eV) around the 
Fermi energy $\epsilon_F$,
where $0.95 \lesssim T_0 \lesssim 1.0$.
The same is true for the results with and without 
charge neutrality.

However, when the geometry is optimized, 
the wide transmission plateaus close to one
are replaced by larger oscillations. Still, at the Fermi energy,
there is usually only a single open channel, which
consists of $s$, $p_z$, and $d_{3z^2-r^2}$ orbitals. 
Sometimes, a small contribution is seen arising from 
a second channel, involving the other $p$ and $d$ orbitals,
as will be discussed below \cite{BrandbygeKobayashiTsukada99}.
The transmission around $\epsilon_F$ varies 
between $0.7 \lesssim T_0 \lesssim 1.0$.
The present method is
known to reproduce experimental conductance histograms 
rather well \cite{DreherPauly05}. 
In particular, the conductance peak somewhat
below $G/G_0=1$ is a very robust feature.

We determine $\epsilon_F$ by charge neutrality in the leads (or bulk). 
Tests with shifting 
its value from this position by $\sim 0.5$ eV 
(which simulates a gate effect) showed 
no significant qualitative effects on the results.
The position of the Fermi energy
with respect to the local electronic structure is still very important.
This is because, besides the elastic transmission, 
the Fermi energy also fixes
the ``Fermi wavevector'', 
which again determines what vibrational modes can be excited.
In the present TB method, the use of bulk parameters and
the charge neutrality procedure 
introduces some uncertainty in relation to this point.

\subsection{Longitudinal and transverse modes}

\begin{figure}[!tb] 
\includegraphics[width=0.65\linewidth,clip=]{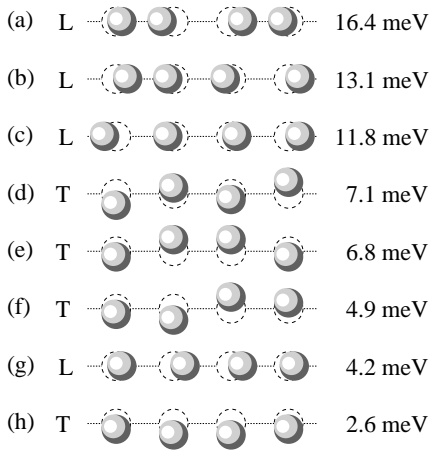}
\caption{A sketch of the vibrational eigenmodes of
a linear four-atom gold wire, with energies corresponding to 
$d=2.62$ {\AA} in geometry A. The longitudinal modes
(L) are all nondegenerate, whereas the transverse modes (T) 
are all doubly degenerate.}
\label{f.fourmodes}
\end{figure}

Let us first discuss the basic observations using a
simple example, namely, a linear chain of four atoms 
in geometry A. A schematic illustration
of the vibrational modes is shown in Fig.\ 
\ref{f.fourmodes} for $d=2.62$ {\AA}.
There are four longitudinal modes and eight transverse modes.
However, due to the fourfold rotation symmetry of the geometry
around the axis of the wire, the transverse modes are
all doubly degenerate. The zero-bias conductance 
is due to two partially open channels. The main contribution
(about 98\% of $G_0$) is due to a channel ($C_1$) formed from 
$s$, $p_z$, and $d_{3z^2-r^2}$ orbitals, which have the symmetry 
of the geometry.
In addition, there is a small 
(less than 1\% of $G_0$) contribution from a second, doubly degenerate 
channel ($C_2$),
which consists of $d_{xz}$, $d_{yz}$, $p_x$, and $p_y$
orbitals, which have a lower symmetry. 
Thanks to the symmetry of the $C_1$ channel, only longitudinal modes
have a finite coupling constant in its subspace 
($\gmat{\lambda}^\alpha_{C_1,C_1}$). 
In the subspace of the $C_2$ channel, 
also the transverse modes have a finite coupling
($\gmat{\lambda}^\alpha_{C_2,C_2}$).
Thus we might expect that also the 
transverse modes give a small signal in the current.

\begin{figure}[!tb] 
\includegraphics[width=0.95\linewidth,clip=]{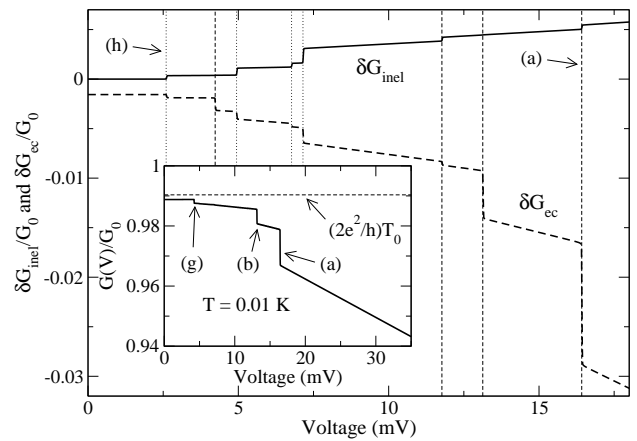}
\caption{
Decomposition of the conductance $G(V)$
into $G_0T_0$, an ``inelastic'' contribution $\delta G_{inel}$, 
and an ``elastic correction'' $\delta G_{ec}$ for a four-atom wire. 
The geometry and the labels (a)-(h) correspond to those of
Fig.\ \ref{f.fourmodes}.
Other parameters
are $T=0.01$ K, $\eta=0.002$ meV.
The solid step-like curve shows $\delta G_{inel}$,
and the dashed one shows $\delta G_{ec}$.
The increases of $\delta G_{inel}$ due to transverse
modes are exactly canceled by decreases in $\delta G_{ec}$.
The inset shows the elastic transmission (dashed line), 
and the total conductance $G(V)$ (solid line). In $G(V)$ only
drops due to longitudinal modes are seen. 
}
\label{f.inel}
\end{figure}

Figure \ref{f.inel} shows an analysis of the
contribution from the different modes to the 
differential conductance $G(V)=\upd I/\upd V$.
We divide this conductance into three parts
according to the three current contributions:
$G(V)=G_0T_0 + \delta G_{inel}(V) + \delta G_{ec}(V)$.
Here $G_0=2e^2/h$, 
$\delta G_{inel}=\upd I_{inel}/\upd V$, 
and $\delta G_{ec}=\upd \delta I_{ec}/\upd V$.
It is seen that $\delta G_{inel}$ gives always
positive contributions to the
conductance steps, while $\delta G_{ec}$
gives negative ones.
As expected, we find that there is a finite step 
in both
$\delta G_{inel}$ and $\delta G_{ec}$ due to 
\emph{all} of the vibrational modes, also 
the transverse ones, although the latter are quite small.
However, the contributions of $\delta G_{inel}$
and $\delta G_{ec}$ for the transverse modes
almost perfectly \emph{cancel} each other,
such that only steps due to the longitudinal 
modes are seen in the total $G(V)$.
This cancellation is apparently due to the exact fourfold 
rotation symmetry,
and the mirror symmetry with respect to the plane cutting
the wire in the middle. In less symmetric geometries
the transverse modes can also give finite contributions to $G(V)$.

In the case of zigzag wires, the distinction between longitudinal 
and transverse modes does not really exist, and all 
modes are always seen as steps in $G(V)$. An example of this
is shown below. Furthermore, if the elastic transmission $T_0$
is very small, then also the transport coefficients 
$T^{ec}_\alpha$ and $T^{ecLR}_\alpha$ tend to be small, 
since they all depend on the matrix 
$\gmatx{G}^r\gmat{\Gamma}_R\gmatx{G}^a\gmat{\Gamma}_L$.
In this way, for example, it may also be possible to 
have large \emph{positive} steps in $G(V)$, but we never see them 
for the charge-neutral gold wires. 
For other materials, the situation may be different.

Thus, we find that the conductance features depend in an intricate way 
on the symmetries of the geometry, the symmetries of the
vibrational modes, the coupling constants, as well as
the symmetries of the
electronic states which are relevant at the Fermi energy.

\subsection{Conductance curves of linear wires}

\begin{figure}[!tb] 
\includegraphics[width=0.95\linewidth,clip=]{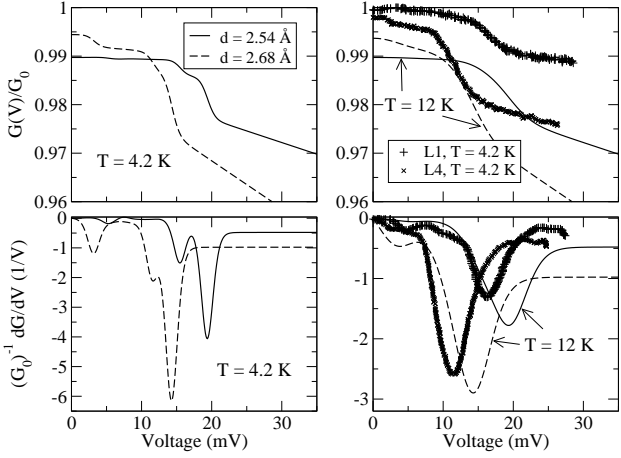}
\caption{Comparison between theory and experiment
for $N_{ch}=4$ in geometry A. 
The solid and dashed curves correspond to theoretical results for
$d=2.54$ {\AA} and $d=2.68$ {\AA}, respectively.
On the left-hand panels $T=4.2$ K, 
whereas on the right-hand panels these
curves have been broadened with a larger temperature 
$T=12$ K; in both cases $\eta=0.02$ meV.
The experimental results L1 ($+$) and L4 ($\times$) correspond to
the notation and results of Fig.\ 1(d) of
Ref.\ \onlinecite{AgraitUntiedt02} with $V>0$.
They are obtained for a 7-atom chain at $T=4.2$ K.
} 
\label{f.COMP1}
\end{figure}
\begin{figure}[!tb] 
\includegraphics[width=0.95\linewidth,clip=]{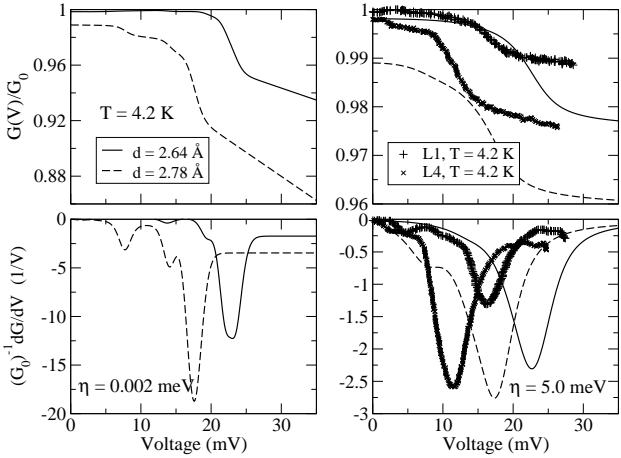}
\caption{Comparison between theory and experiment
for $N_{ch}=11$ in geometry B. 
All results are at $T=4.2$ K.
The solid and dashed curves correspond to theoretical results for
$d=2.64$ {\AA} and $d=2.78$ {\AA}, respectively.
On the left-hand panels $\eta=0.002$ meV, whereas on the
right-hand panels the 
curves have been broadened with a bath-coupling $\eta=5.0$ meV.
The experimental results 
L1 ($+$) and L4 ($\times$) are as in Fig.\ \ref{f.COMP1}.
}
\label{f.COMP2}
\end{figure}
Here we discuss in more detail, how our conductance-voltage curves
for linear wires look like.
Figure \ref{f.COMP1} shows an example
for geometry A with a wire of $N_{ch}=4$ atoms
while Fig.\ \ref{f.COMP2} is for a wire of $N_{ch}=11$ atoms in geometry B.
In both figures, the left-hand panels are calculated at $T=4.2$ K and
with a small $\eta$, such that they are more or less in the externally 
undamped regime.
The right-hand panels show two examples of the experimental
results for a wire of approximately 7 atoms 
taken at the temperature $T=4.2$ K \cite{AgraitUntiedt02}. 
Comparing these to the theoretical curves on the left-hand side, 
one immediately notices 
that if the conductance drop is to be due to a single mode, then the 
$\sim 5$ meV width of the peak
in the experimental $\upd G/\upd V$ cannot be explained by temperature 
alone \cite{Frederiksen04a}.
On the other hand, the energy distance between the vibrational
modes is rather large $\gtrsim k_BT$, so that at $T=4.2$ K, a peak
consisting of
several sub-peaks can usually be easily recognized.
For example, the highest-frequency peak in Fig.\ \ref{f.COMP2}
actually consists of two peaks, and it is still not wide enough.
Also in Fig.\ \ref{f.COMP1} 
there are at least three separate peaks visible.

Thus we conclude
that in the experiment there are probably other broadening mechanisms
at play besides temperature. 
In the right panels of Fig.\ \ref{f.COMP1} we compare the experiment
with a theoretical result broadened by a higher temperature,
while in the right panels of Fig.\ \ref{f.COMP2} we use the 
parameter $\eta$ to broaden the peaks. 
In the latter, the system is already
in the externally damped regime, with very little local heating: 
in addition to the broadening, the damping is signified by a smaller
slope after the drop. 
In either case, the peaks due to individual modes
are smoothed out to form a single one, with a width comparable to 
that seen in experiments.
In this way, it is possible to obtain a rather good
quantitative correspondence between theory and experiment.

\begin{figure}[!tb] 
\includegraphics[width=0.99\linewidth,clip=]{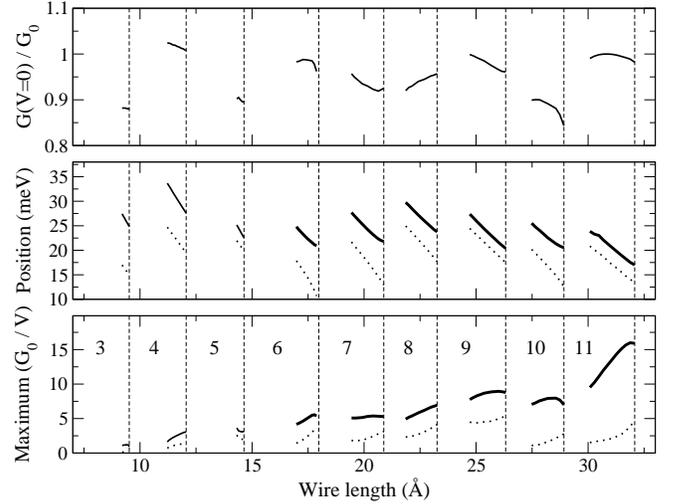}
\caption{
Zero-bias conductances,
voltage positions $V_{ph}$, and the local maxima of $|\upd G/\upd V|$
for different wire lengths $L_{ch}$ and atom numbers $N_{ch}$,
indicated by the numbers. 
The results are obtained for linear wires in geometry B, with
$\eta=0.02$ meV.
The dashed lines show the limits beyond which the wire breaks
upon further stretching.
On the left side of each curve, the wire has a non-linear form,
and the peak structure may be different (see text).
The solid lines correspond to the main peaks 
and the dotted lines to the second-largest
ones.
Thickening of the lines indicates that the peak 
consists of two close-lying modes. 
} 
\label{f.peaks}
\end{figure}
We have also studied systematically how the main features of the
conductance-voltage curves vary when linear chains with atom numbers
$3\leq N_{ch}\leq11$ are stretched. The results for $T=4.2$ K and
a small $\eta$ are plotted
in Fig.\ \ref{f.peaks}.
Here we show the zero-bias conductance $G(V=0)$,
the voltage positions $V_{ph}$ of the main peaks in $\upd G/\upd V$,
and the corresponding local maximum values of $|\upd G/\upd V|$.
The peak height is not the same as the peak area which is plotted in
Ref.\ \onlinecite{AgraitUntiedt02}, but is roughly
proportional to it, since the width of the peaks is 
$\sim k_BT$, where the temperature is $T=4.2$ K. 

In this figure, many features can be recognized.
The $G(V=0)$ values fluctuate between $0.85$ and $1.0$,
but there is no clear ``parity effect'' --- or at least the
effect appears to reverse its direction after $N_{ch}=8$.
The present method most likely does not describe 
such parity effects correctly. However, also different
\emph{ab initio} approaches are known to give 
conflicting results \cite{SmitUntiedt03}.
In some calculations, the transmission has been 
found to oscillate also with the stretching of a wire with 
fixed $N_{ch}$ \cite{LeeBrandbyge04}.

The positions of the peaks move to lower voltages when the 
wire is stretched, 
as expected from the ``softening'' of the bonds, 
and the resulting decrease in the vibrational frequencies.
There is also a clear trend toward lower frequencies with
increasing $N_{ch}$. 
These findings are similar to what is seen in the experiment 
\cite{AgraitUntiedt02}.
Also similarly, the peak heights increase with stretching,
and with increasing $N_{ch}$, although the increase with 
wire length $L_{ch}$
is not obviously linear as for the simple chain models of 
Sec.\ \ref{s.simple} \cite{Frederiksen04b}.
There is also a correlation between the
zero-bias conductances and the peak heights: 
when the conductance is low, also the stretching
behavior is rather anomalous (in particular in the 
cases of 5 and 10 atoms.)

The most visible difference between the 
results of Fig.\ \ref{f.peaks}
and the experiment is that we consistently see 
signatures of several vibrational modes:
typically there are two peaks visible in $\upd G/\upd V$. 
However, the higher peak is always at a larger voltage
and, as the number of atoms $N_{ch}$ grows, the secondary peaks become 
less and less significant. For example, for $11$ atoms there is 
essentially only a single peak visible (see Fig.\ \ref{f.COMP2}).
This, however, is due to two close-lying modes: the 
highest-frequency ``ABL'' mode, and the one next to it in frequency.
As explained above, this discrepancy of several peaks can be 
corrected by increasing the parameter $\eta$.
Note also that this behavior was already present in the
chain model of Sec.\ \ref{s.simple}.

We also see that the largest 
conductance drops are systematically at too high 
voltages compared to experimental values $V_{ph}\approx10-20$ mV
for a 7-atom chain. 
This is not surprising, given the simplified way in which we 
compute the vibrational modes.
The frequencies of the vibrational modes might be lowered, 
if we also allowed for the motion of atoms outside of the chain.
In other words, it is possible that the lead coupling, done 
in a proper way, would lead to a ``redshift'' of the frequencies.
As noted above, the electron-vibration coupling gives such a 
redshift \cite{MiiTikhodeevUeba03}, but this 
effect may be too small to explain the discrepancy.

Although the steps in conductance are almost always downward
when $G(V=0)$ is close to $G_0$, sometimes also weak increases in
the conductance at low voltage can be seen. These appear to be
related to the longitudinal ``center-of-mass'' mode.
For linear chains, we do not find any significant contribution
from transverse modes, as explained previously. 

\subsection{Zigzag wires}

\begin{figure}[!tb] 
\includegraphics[width=0.95\linewidth,clip=]{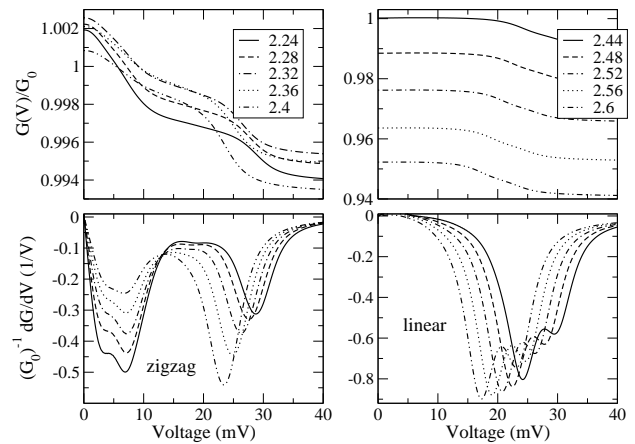}
\caption{The difference between zigzag and linear 
wires of 6 atoms in geometry A at $T=4.2$ K and with $\eta=5.0$ meV.
The numbers in the legend indicate the parameter $d/$\AA, which 
describes the amount of stretching.
In the left panels the wire has a zigzag character,
while in the right panels it is linear.
In case of the zigzag-wire, practically all of the vibrational
modes contribute to the observed steps, but they
form a clear double-step structure.
In the linear wire, the elastic transmission varies
a lot under stretching in this example.
} 
\label{f.ZIGZAG6}
\end{figure}
We have also studied briefly the zigzag chains shown in Fig.\ \ref{f.lcr}.
Figure \ref{f.ZIGZAG6} shows a comparison between 
the signatures of a zigzag chain and a linear chain
in the conductance-voltage curves $G(V)$.
In this example, a chain of $N_{ch}=6$ atoms in geometry A, and
a large broadening of the vibrational
modes ($\eta=5.0$ meV) was used.
However, very similar results were found for various different
atom numbers. 
For small values of $d$, the wire has a zigzag character,
but at $d\approx2.42$ {\AA}, the wire becomes linear.

For the zigzag chain, there are two well-separated (series of)
conductance drops. The one at low bias is higher
for small $d$'s. As the 
chain is stretched, the height of the lower-frequency step decreases,
while that of the higher-frequency one increases.
For the linear chain, there is essentially only one
high-voltage drop, which is actually due to two 
different longitudinal modes.
In the case of a zigzag wire, 
the vibrational modes have complicated symmetries, and
practically all of them are contributing to the current
steps. However, here their identities are completely smeared out
due to the broadening. 
This calculation provides a clear prediction of how 
the zigzag-to-linear transition may be seen in experiments.

\section{Conclusions and discussion} \label{s.conc}

We have studied the onset of dissipation by excitation
of vibrational modes in atomic gold wires, using a
tight-binding model. 
We have studied in a systematic way how the
stretching of wires with different atomic numbers
affects the conductance steps, and find a reasonable 
agreement with experiments.
Previously such a study has only been done within 
simple chain models \cite{Frederiksen04b}.
We have also considered two different geometries, which 
yield qualitatively similar results.
Finally, we studied the conductance signatures
of zigzag-type wires, and predict a double-step
structure in contrast to the single step of linear chains.

Our results for the linear chains agree rather well with 
experiments and previous \emph{ab initio} calculations,
apart from the incomplete ``mode selectivity''.
In this context, we have pointed out the importance of taking into 
account the broadening of vibrational modes due to
their coupling to the leads, especially in the
limit where the vibrational mode distribution 
is assumed to be strongly damped. 
We derived equations in the
wide-band limit, which take this into account in a
phenomenological manner. 
The wide-band limit combined with
the lowest-order perturbation approach (presented in 
Sec.\ \ref{s.wbl} and Appendix \ref{s.form}) appears to 
provide a sufficiently
good description of the phenomenology
of electron-vibration interaction in atomic wires. 
To make further progress, more detailed calculations of the
lead-coupling of the vibrational modes are needed.

As explained above, 
the condition which determines what modes yield 
conductance drops is essentially that of momentum
conservation.
We find numerically that the momentum-conservation idea
works well also for finite wires with 10 atoms or more.
However, in shorter wires, the 
electronic structure is very complicated, and 
the connection between symmetries of excited modes and those
of the electronic states is hard to analyze.

Even if there is approximate momentum conservation,
there is no fundamental reason why only a single mode would be excited, 
or that the highest-frequency (or other ABL) modes 
should necessarily be involved.
This remains true also for
very long wires, since at the same time when the momentum conservation 
becomes more and more accurately satisfied, also the energy density 
(and hence momentum-space density) 
of the modes increases. It is thus more likely
that a large ``wave packet'' of several nearby phonon modes is always 
excited.
However, the charge neutrality of the wire indeed appears 
to be favoring the highest-frequency modes,
just as predicted by the simple chain models 
discussed in Sec.\ \ref{s.simple}.
Nevertheless, it is important to note that small differences
in details of the geometry or the implementation
may affect numerical values of results significantly. 
The band structure
of an infinite, linear gold wire is already quite complicated 
\cite{BrandbygeKobayashiTsukada99,SanchezPortal99}, such that a small 
error in the Fermi energy, and hence of the Fermi wavevector, 
will immediately shift the resonance to slightly different 
vibrational modes.

Studying different materials (Pt and Ir) with this same method 
would be straightforward in principle. However, it seems that
the parameters available for these materials are not very good for 
geometry optimizations of the wires. This is because the overlap matrices 
easily lose their positive definiteness 
when the validity range of the parametrization is exceeded
(with the gold parameters \cite{website}, such problems never appeared).
Thus, as already implied in 
Ref.\ \onlinecite{BrandbygeKobayashiTsukada99},
one should probably use more general \emph{ab initio} methods,
or at least parameters which have been specifically fit
to work for chain geometries with a large span of interatomic distances.
Compared to \emph{ab initio} methods, TB still has the clear 
advantage of being computationally efficient.

\acknowledgments

It is a pleasure to acknowledge useful discussions with N. Agra\"it, M.
Brandbyge, T. Frederiksen, A. Levy-Yeyati, A. Martin-Rodero and G. Rubio.
This work was financially supported by the Helmholtz Gemeinschaft within
the Nachwuchsgruppe program (contract VH-NG-029) and by the DFG within the
Center for Functional Nanostructures.

\appendix

\section{Calculation of the coupling matrix elements} \label{s.cccal}

Here we describe our method 
of computing the matrix elements
$M_{ij}^{k\mu}=\langle i|\nabla_{k\mu}H_{}|j\rangle$,
where the derivative is with respect to the 
components $\mu=x,y,z$ of
the ionic coordinates $\vec R_k$. 
In an implementation making use of TB
parametrization, one has no direct access 
to either the basis states or $|i\rangle$, or the 
Hamiltonian $H_{}$ --- only the representation 
$H_{ij}=\langle i|H_{}|j\rangle$ and the overlap matrix
$S_{ij}=\langle i|j\rangle$ as a function of $\vec R_k$ 
are known.  
Thus, the matrix elements must be calculated more indirectly.
Below we sketch a derivation, which follows, in some sense, 
the ideas of Ref.\ \onlinecite{HeadGordonTully91}.
The derivation is not exact, as we only consider an isolated 
central cluster.

Let $|i\rangle,|j\rangle,\ldots$ denote the atomic orbital (AO) basis states, 
and $|\alpha\rangle,|\beta\rangle,\ldots$ the electronic eigenstates 
(molecular orbitals, MO) of the central-cluster electronic Hamiltonian $H$. 
The eigenstates satisfy
\begin{equation}\begin{split} \label{e.mos}
H_{}|\alpha\rangle&=\epsilon_\alpha|\alpha\rangle, \\
\langle \alpha|\beta\rangle &= \delta_{\alpha\beta}. \\
\end{split}\end{equation}
If the expansion of the
MO's in the AO basis is denoted 
$|\alpha\rangle=\sum_i |i\rangle C_{i\alpha}$, where 
$C_{i\alpha}=(\cvec_\alpha)_i$ then
we have the matrix equations
\begin{equation}\begin{split} \label{e.aobas}
\gmat H\cvec_\alpha &= \gmat S\cvec_\alpha\epsilon_\alpha \\
\cvec^\dagger_\alpha\gmat S\cvec_\beta &= \delta_{\alpha\beta}.
\end{split}\end{equation}
Let us also note the form of the completeness relation of the MO's
$\sum_\alpha|\alpha\rangle\langle\alpha|=1$ in the AO basis:
$\sum_\alpha\cvec_\alpha\cvec_\alpha^\dagger=\gmat S^{-1}$, while
for the AO's themselves 
$\sum_{ij}|i\rangle(\gmat S^{-1})_{ij}\langle j|=1$.
Since the basis states also move with the ions, we should write 
more carefully
$|i(Q)\rangle$, $|\alpha(Q)\rangle$, $H_{}(Q)$, $H_{ij}(Q)$ etc.,
where $Q$ is a shorthand for the ionic displacements
$\vec Q_k=\vec R_k-\vec R^{(0)}_k$.
The point in the electron-vibration coupling is that moving an 
ion will induce a perturbation of the form
\begin{equation}\begin{split}
\gmat H &\rightarrow \gmat H+\gmat H'Q  \\
\gmat S &\rightarrow \gmat S+\gmat S'Q
\end{split}\end{equation}
and one can calculate its effect on 
the state vectors $\cvec_\alpha$ and eigenenergies
$\epsilon_\alpha$ by means of simple first-order perturbation 
theory. Here ``~$'$~'' denotes a derivative with respect to $Q$.

Let us expand the basis states as 
$|i(Q)\rangle\approx|i\rangle+|i'\rangle Q$,
where $|i\rangle$ now denote the unperturbed basis.
The matrix element which we are looking for is really the
quantity 
\begin{equation}\begin{split}
M_{ij}=\langle i|H_{}'(Q)|j\rangle\big|_{Q=0}.
\end{split}\end{equation}
This may be obtained by considering the expansion of
$\langle i|H_{}(Q)|j\rangle$ in $Q$, since
\begin{equation}\begin{split}
\langle i|H_{}(Q)|j\rangle
=
H_{ij}(0)
+\langle i|H_{}'(Q)|j\rangle\big|_{Q=0}Q.
\end{split}\end{equation}
By inserting $1=\sum_{\alpha}|\alpha(Q)\rangle\langle\alpha(Q)|$ one
easily finds
\begin{equation}\begin{split}
\langle i|H_{}(Q)|j\rangle
=\sum_\alpha\langle i|\alpha(Q)\rangle
\epsilon_\alpha(Q)\langle\alpha(Q)|j\rangle.
\end{split}\end{equation}
Then, by inserting 
$|\alpha(Q)\rangle=\sum_i|i(Q)\rangle C_{i\alpha}(Q)$,
expanding 
$\epsilon_{\alpha}(Q)\approx \epsilon_{\alpha}+\epsilon'_{\alpha}Q$,
$C_{i\alpha}(Q)\approx C_{i\alpha}+C'_{i\alpha}Q$,
and equating terms linear in $Q$
\begin{equation}\begin{split} \label{e.matel2}
M_{ij}
&=
\sum_\alpha\sum_{kl}\big\{
S_{ik}C_{k\alpha}\epsilon'_\alpha C_{l\alpha}^*S_{lj} \\
&+S^{'(2)}_{ik}C_{k\alpha}\epsilon_\alpha C_{l\alpha}^*S_{lj} 
+S_{ik}C_{k\alpha}\epsilon_\alpha C_{l\alpha}^*S^{'(1)}_{lj} \\
&+S_{ik}C'_{k\alpha}\epsilon_\alpha C_{l\alpha}^*S_{lj}
+S_{ik}C_{k\alpha}\epsilon_\alpha C_{l\alpha}^{'*}S_{lj}
\big\}.
\end{split}\end{equation}
Here we introduced the one-sided overlap derivatives 
$[\gmat S^{'(1)}]_{ij}=\langle i'|j\rangle$,
$[\gmat S^{'(2)}]_{ij}=\langle i|j'\rangle$,
which satisfy
$\gmat S' = \gmat S^{'(1)}+\gmat S^{'(2)}$, and 
$\gmat S^{'(2)}=[\gmat S^{'(1)}]^\dagger$ \cite{HeadGordonTully91}. 

The result of Eq.\ (\ref{e.matel2}) may be
simplified considerably by inserting the expressions 
for $\epsilon_\alpha'$ and $C'_{i\alpha}$, which are easily derived.
First, if we denote
\begin{equation}\begin{split}
M_{\alpha\beta}&\equiv
\sum_{kl}C_{k\alpha}^*M_{ij}C_{l\beta}, \quad
H_{\alpha\beta}^{'}\equiv
\sum_{kl}C_{k\alpha}^*H_{ij}^{'}C_{l\beta}, \\
S_{\alpha\beta}^{'(1,2)}&\equiv
\sum_{kl}C_{k\alpha}^*S_{ij}^{'(1,2)}C_{l\beta}
\end{split}\end{equation}
and so on, one may show that
\begin{equation}\begin{split}
M_{\alpha\beta}=H_{\alpha\beta}'
-S^{'(1)}_{\alpha\beta}\epsilon_\beta
-\epsilon_\alpha S^{'(2)}_{\alpha\beta}.
\end{split}\end{equation}
Then, using the completeness relations
$\sum_{\alpha}\sum_{k}S_{ik}C_{k\alpha}C_{j\alpha}^*
=\sum_{\alpha}\sum_{k}C_{i\alpha}C_{k\alpha}^*S_{kj}=\delta_{ij}$
one may transform back to the AO basis:
\begin{equation}\begin{split}
M_{ij}&= H'_{ij} - \sum_\alpha\sum_{kl}\big\{ \\
&S^{'(1)}_{ik}C_{k\alpha}\epsilon_\alpha C_{l\alpha}^*S_{lj} 
+S_{ik}C_{k\alpha}\epsilon_\alpha C_{l\alpha}^*S^{'(2)}_{lj}
\big\}.
\end{split}\end{equation}
This is the final expression for the matrix elements.
The overlap corrections due to the nonorthogonal basis
often turn out to be rather small in practice, an 
the orthogonal result $\gmat{M}=\gmat{H}'$ is
a reasonable first approximation.
The presence of the corrections tends to make 
the conductance steps slightly larger.

\section{Population analysis and
derivation of the current formula} \label{s.cder}

Although the expression for the current is
well known, its derivation in the nonorthogonal basis
is not entirely trivial \cite{XueDattaRatner02}.
The current is the time derivative 
of charge transported from $L$
to $C$ and on to $R$. However, 
while the total charge of the 
full system is well defined,
the partial
charges of the distinct regions are
not --- instead, there
are different ways of doing 
``population analysis''.
To be more self-contained, we present
here a compact discussion of these
issues.

Let us consider, for simplicity, a single orbital per atomic site $i$.
The total charge (particle number, actually) is given by
\begin{equation}\begin{split}
Q=\langle\hat Q\rangle=\sum_{j,k} P_{jk}S_{kj},
\end{split}\end{equation}
where
$\op Q=\sum_{j,k} d_{k}^\dagger S_{kj}d_{j}$
and we define the density matrix 
$P_{jk}=\langle d_{k}^\dagger d_{j}\rangle$.
Partial charges may be introduced for example with the
following ``Mulliken'' population analysis \cite{SzaboOstlund}
\begin{equation}\begin{split}
Q&=Q_L+Q_C+Q_R
=\sum_{j\in L,k} P_{jk}S_{kj} \\
&+\sum_{j\in C,k} P_{jk}S_{kj} 
+\sum_{j\in R,k} P_{jk}S_{kj}.
\end{split}\end{equation}
One may also define the quantities
\begin{equation}\begin{split}
Q_{\Omega}' &= \langle\op Q_{\Omega}'\rangle
=\sum_{j,k\in \Omega} P_{jk}S_{kj}, \\ 
\end{split}\end{equation}
where 
${\op Q}'_\Omega=\sum_{j,k\in\Omega} d_{k}^\dagger S_{kj}d_{j}$,
and $\Omega=L,C,R,C+R,C+L$. Here
$C+L$ ($C+R$) refers to the combined system involving
$C$ and $L$ ($C$ and $R$) regions.
The charges $Q'_{L,C,R}$ are good approximations to 
$Q_{L,C,R}$ if the regions are all large, 
since the corrections are proportional to the 
surface area of the interfaces.

What we want is the particle current $\op J$ through a boundary
between two regions of space, $L$ and $C$, for example.
We expect it to satisfy 
\begin{equation}
2\op J=\frac{\partial}{\partial t}({\op Q}'_L-{\op Q}'_{C+R}).
\end{equation}
Let us rewrite 
${\op Q}'_\Omega={\op N}_\Omega+\sum_{j,k\in\Omega,j\neq k} d_{k}^\dagger S_{kj}d_{j}$,
where
$\op N_{\Omega}=\sum_{j\in \Omega}\op N_j$
and $\op N_j=d^\dagger_{j}d_j$ for $\Omega=L,C+R$.
Now, in the Heisenberg picture the $d_j$ operators
satisfy the equations of motion $\iu\hbar\dot d_j=[d_j,\op H_{e}]$ 
and hence $\iu\hbar\frac{\partial}{\partial t}\sum_{k}S_{jk}d_k=\sum_{k}H_{jk}d_k$.
Using ideas similar to Ref.\ \onlinecite{EmberlyKirczenow98}, we find
\begin{equation}\begin{split}
\frac{\partial}{\partial t}{\op N}_j
= d_j^\dagger \sum_{k\neq j}
\frac{1}{\iu\hbar}\left(H_{jk}-S_{jk}\iu\hbar\frac{\partial}{\partial t}
\right)d_k + \textrm{h.c.}
\end{split}\end{equation}
Next, considering the following quantity, we notice that
all hopping contributions except
$\gmat{H}_{LC}$ and $\gmat{H}_{LR}$ cancel: 
\begin{equation}\begin{split}
\frac{\partial}{\partial t} &
\left(\op N_L-\op N_{C+R}\right) \\
&= \sum_{j\in L,k\in C+R}
d_j^\dagger 
\frac{1}{\iu\hbar}\left(H_{jk}-S_{jk}\iu\hbar\frac{\partial}{\partial t}
\right)d_k + \textrm{h.c.} \\
&- \sum_{j\in C+R, k\in L}
d_j^\dagger 
\frac{1}{\iu\hbar}\left(H_{jk}-S_{jk}\iu\hbar\frac{\partial}{\partial t}
\right)d_k + \textrm{h.c.} \\
&- \sum_{j,k\in L,k\neq j}
d_j^\dagger 
S_{jk}\frac{\partial}{\partial t}d_k + 
\textrm{h.c.} \\
&+ \sum_{j,k\in C+R,k\neq j}
d_j^\dagger 
S_{jk}\frac{\partial}{\partial t}d_k + 
\textrm{h.c.} \\
\end{split}\end{equation}
When the last two overlap terms
are moved to the left-hand side, it 
becomes exactly what we called $2\op J$.
Thus the expectation value is
\begin{equation}\begin{split}
2J &= 2\langle \op J \rangle \\
&= \sum_{j\in L,k\in C+R}
\frac{1}{\iu\hbar}\left(H_{jk}-S_{jk}\iu\hbar\frac{\partial}{\partial t}\right)
\langle d_j^\dagger(t') d_k(t) \rangle\bigg|_{t'=t} \\
&- \sum_{j\in C+R,k\in L}
\frac{1}{\iu\hbar}\left(H_{jk}-S_{jk}\iu\hbar\frac{\partial}{\partial t}\right)
\langle d_j^\dagger(t') d_k(t)\rangle\bigg|_{t'=t} \\
&+\textrm{c.c.}
\end{split}\end{equation}
Now, assuming that $\gmat{H}_{LR}=\gmat{S}_{LR}=0$
and using $\iu\langle d_j^\dagger(t') d_k(t) \rangle=G^{+-}_{kj}(t,t')$, we 
finally have
\begin{equation}\begin{split}
J(t) &= -\frac{1}{\hbar}\re\Tr\bigg[
\left(\gmat{H}_{LC}-\gmat{S}_{LC}\iu\hbar\frac{\partial}{\partial t}\right)
\gmat{G}^{+-}_{CL}(t,t') \\
&- 
\left(\gmat{H}_{CL}-\gmat{S}_{CL}\iu\hbar\frac{\partial}{\partial t}\right)
\gmat{G}^{+-}_{LC}(t,t')
\bigg]\bigg|_{t'=t}.
\end{split}\end{equation}
The charge current is then obtained as $I=-2eJ$, where the factor
$2$ is the spin degeneracy.
An analogous derivation may be carried out for the current over the 
$C$-$R$ boundary.

In stationary state
all the propagators only depend on the time difference $t-t'$,
and this result may be Fourier transformed into an energy representation,
where one has to replace
$\iu\hbar\partial/\partial t\rightarrow \epsilon$.

\section{NEGF formalism: technical details} \label{s.negf}

Our notation differs slightly from what is the standard.
In particular, our functions $\gmat{G}^{\pm\mp}$ are equal to 
the $\gmat{G}^{\lessgtr}$ functions of 
Refs.\ \cite{Mahan,Frederiksen04a}, for example.
Let us write the definitions for the most important
electron propagators
\begin{equation}\begin{split} \label{e.gdef}
G_{ij}^r(t,t')&=-\iu\langle \{d_i(t),d_j^\dagger(t')\}\rangle\theta(t-t') \\
G_{ij}^a(t,t')&=\iu\langle \{d_i(t),d_j^\dagger(t')\}\rangle\theta(t'-t) \\
G_{ij}^{+-}(t,t')&=\iu\langle d_j^\dagger(t')d_i(t)\rangle \\
G_{ij}^{-+}(t,t')&=-\iu\langle d_i(t)d_j^\dagger(t')\rangle. \\
\end{split}\end{equation}
Similar expressions hold for phonons, but with the replacement
$\{\cdot,\cdot\}\rightarrow [\cdot,\cdot]$.
Below, all Green functions appear Fourier transformed with respect to
$t-t'$ into an energy representation.

\subsection{Electron propagators and self-energies}

All expressions for the 
relevant electron Green functions follow from the 
``Dyson'' and ``Keldysh'' equations
\begin{equation}\begin{split}
\gmat{G}^r&=[(\gmat{g}^r)^{-1}-\gmat{\Sigma}^r]^{-1} \\
\gmat{G}^{\pm\mp} &= (1+\gmat{G}^r\gmat{\Sigma}^r)
\gmat{g}^{\pm\mp}(1+\gmat{\Sigma}^a \gmat{G}^a)
-\gmat{G}^r\gmat{\Sigma}^{\pm\mp}\gmat{G}^a,
\end{split}\end{equation}
where the upper or lower signs can be chosen.
Here the $\gmat{g}$ denote the Green functions for
an uncoupled central part in the absence of electron-vibration
interactions
and $\gmat{\Sigma}^{r,\pm\mp}$ are the sums of all self-energies 
containing the effects of both (see below). 
The uncoupled functions are
\begin{equation}\begin{split}
\gmat{g}^r&=[\epsilon\gmat{S}_{CC}+\iu\gmat{\gamma}_c/2-\gmat{H}_{CC}]^{-1}\\
\gmat{g}^{+-} &= -f^{}_{C}(\gmat{g}^r-\gmat{g}^a)  \\
\gmat{g}^{-+} &= -(f^{}_{C}-1)(\gmat{g}^r-\gmat{g}^a).  \\
\end{split}\end{equation}
Here $f^{}_{C}$ is the equilibrium Fermi distribution.
Note that
$\gmat{\gamma}_{C}=\iu[(\gmat{g}^a_{})^{-1}-(\gmat{g}^r_{})^{-1}]$,
which is an infinitesimal quantity. 

The functions where the lead coupling is taken into account but
electron-vibration coupling is still neglected are given by
\begin{equation}\begin{split}
\gmatx{G}^r & = [\epsilon\gmat{S}_{CC}-\gmat{H}_{CC}
-\gmat{\Sigma}^r_L -\gmat{\Sigma}^r_R  ]^{-1} \\
\gmatx{G}^{+-} & = -\gmat{G}^r(\gmat{\Sigma}_L^{+-}+\gmat{\Sigma}_R^{+-}
-\iu\gmat{\gamma}_{C}f_{C}^{})\gmat{G}^a \\
\gmatx{G}^{-+} & = -\gmat{G}^r[\gmat{\Sigma}_L^{-+}+\gmat{\Sigma}_R^{-+} 
-\iu\gmat{\gamma}_{C}(f_{C}^{}-1)]\gmat{G}^a. \\
\end{split}\end{equation}
and $\gmatx{G}^a=(\gmatx{G}^r)^\dagger$.
Here the lead self-energies and lead Green functions for the $L$ side are 
given by
\begin{equation}\begin{split}
\gmat{\Sigma}^r_L & = \gmat{t}_{CL}\gmat{g}^r_{LL}\gmat{t}_{LC} \\
\gmat{g}^r_{LL} & = [(\epsilon+\iu\gamma_L/2)
\gmat{S}_{LL}-\gmat{H}_{LL}]^{-1} \\
\gmat{\Gamma}_L & = \iu (\gmat{\Sigma}^r_L - \gmat{\Sigma}^a_L) \\
\gmat{\Sigma}_L^{+-} & = \gmat{t}_{CL}\gmat{g}^{+-}_{LL}\gmat{t}_{LC} 
= -\iu \gmat{\Gamma}_L f_L \\
\gmat{\Sigma}_L^{-+} & = \gmat{t}_{CL}\gmat{g}^{-+}_{LL}\gmat{t}_{LC}
= -\iu \gmat{\Gamma}_L (f_L-1), \\
\end{split}\end{equation}
where 
$\gmat{t}_{LC}=\gmat{H}_{LC}-\epsilon\gmat{S}_{LC}$ and so on,
with similar expressions for $R$ side.
The infinitesimal $\gmat{\gamma}_{C}$ is only needed for recovering 
the correct result in 
the limit where the self-energies 
are taken to zero --- it may be neglected here.
The parameters $\gamma_{L,R}$ are positive infinitesimals, 
which, however, can be used to introduce a finite broadening
of the lead eigenstates.

These are enough for calculating the elastic current in the absence of
electron-vibration coupling.
The full Green functions including the effects of this coupling are
\begin{equation}\begin{split}
\gmat{G}^r & = [\epsilon\gmat{S}_{CC}-\gmat{H}_{CC}
-\gmat{\Sigma}^r_L -\gmat{\Sigma}^r_R -\gmat{\Sigma}^r_{e-vib} ]^{-1} \\
\gmat{G}^{+-} & = -\gmat{G}^r(\gmat{\Sigma}_L^{+-}+\gmat{\Sigma}_R^{+-}
+\gmat{\Sigma}^{+-}_{e-vib}-\iu\gmat{\gamma}_{C}f_{C}^{})\gmat{G}^a \\
\gmat{G}^{-+} & = -\gmat{G}^r[\gmat{\Sigma}_L^{-+}+\gmat{\Sigma}_R^{-+} 
+\gmat{\Sigma}^{-+}_{e-vib}-\iu\gmat{\gamma}_{C}(f_{C}^{}-1)]\gmat{G}^a \\
\end{split}\end{equation}
and $\gmat{G}^a=(\gmat{G}^r)^\dagger$.
To second order in the coupling constant $\gmat{\lambda}^\alpha$,
the electron-phonon self-energies are 
\begin{equation}\begin{split} \label{e.fullsigmas}
\gmat{\Sigma}^{\pm\mp}_{e-vib}(\epsilon) = &
-\iu \sum_\alpha \int\frac{\upd\omega_1}{2\pi}
D_{\alpha}^{\pm\mp}(\omega_1)[\gmat{\lambda}^\alpha
\gmat{G}^{\pm\mp}(\epsilon-\omega_1)\gmat{\lambda}^\alpha] \\
\gmat{\Sigma}^{r}_{e-vib}(\epsilon) & = \iu\sum_\alpha
\int\frac{\upd\omega_1}{2\pi}
\big\{
D^{\pm\mp}_\alpha(\omega_1)[\gmat{\lambda}^\alpha
\gmat{G}^{r}(\epsilon-\omega_1)\gmat{\lambda}^\alpha] \\
&+D^{r}_\alpha(\omega_1)[\gmat{\lambda}^\alpha
\gmat{G}^{\mp\pm}(\epsilon-\omega_1)\gmat{\lambda}^\alpha] \\
&-\gmat{\lambda}^\alpha\Tr[\gmat{G}^{+-}(\omega_1)\gmat{\lambda}^\alpha]D^r_\alpha(0)
\big\} \\
\gmat{\Sigma}^{a}_{e-vib}(\epsilon) & = \iu\sum_\alpha
\int\frac{\upd\omega_1}{2\pi}
\big\{
D^{\pm\mp}_\alpha(\omega_1)[\gmat{\lambda}^\alpha
\gmat{G}^{a}(\epsilon-\omega_1)\gmat{\lambda}^\alpha] \\
&+D^{a}_\alpha(\omega_1)[\gmat{\lambda}^\alpha
\gmat{G}^{\mp\pm}(\epsilon-\omega_1)\gmat{\lambda}^\alpha] \\
&-\gmat{\lambda}^\alpha\Tr[\gmat{G}^{+-}(\omega_1)\gmat{\lambda}^\alpha]D^r_\alpha(0)
\big\}, \\
\end{split}\end{equation}
where the upper or lower signs are chosen.

\subsection{Phonon propagators and self-energies}

By our definition of the phonon propagators, 
the unperturbed ones (those in the absence of a lead coupling and
electron-vibration coupling) are given by
\begin{equation}\begin{split} \label{e.bared}
d^r_\alpha(\epsilon)&=\frac{1}{\epsilon-\epsilon_\alpha+\iu\eta/2}
-\frac{1}{\epsilon+\epsilon_\alpha+\iu\eta/2} \\
&=
\frac{2\epsilon_\alpha}{\epsilon^2-\epsilon_\alpha^2
+\iu\eta\epsilon-\eta^2/4}\\
d_\alpha^{+-}(\epsilon) & = -2\pi\iu n(\epsilon)\rho_\alpha(\epsilon) \\
d_\alpha^{-+}(\epsilon) & = -2\pi\iu [n(\epsilon)+1]\rho_\alpha(\epsilon).
\end{split}\end{equation}
Here $\epsilon_\alpha=\hbar\omega_\alpha$ are the bare
vibrational energies,
$n(\epsilon)$ is the Bose distribution function.
The quantity $\rho_\alpha=-\im d^r_\alpha/\pi$ is
the bare phonon density of states (DOS),
given by Eq.\ (\ref{e.apprphdos}).
This becomes
$\rho_\alpha(\epsilon)=\delta(\epsilon-\epsilon_\alpha)$
$-\delta(\epsilon+\epsilon_\alpha)$, 
as $\eta\rightarrow0+$.
Note that the Green functions satisfy
$\eta=\iu[(d^a)^{-1}-(d^r)^{-1}]$.

Now, to be symmetric with the discussion for the electron propagators, 
the next step should be the introduction propagators ``$\td{D}$''
which contain lead self-energies.
For the proper calculation of the lead self-energies, 
we should probably change to an atomic-displacement basis.
Instead of doing this, we model the 
lead coupling by giving finite values to the infinitesimal 
quantity $\eta$, which will broaden the phonon density of 
states \cite{GalperinRatnerNitzan04}.

The full phonon propagators appearing in the electron-phonon self-energies 
are obtained from the ``Dyson'' and ``Keldysh'' 
equations for phonons
\begin{equation}\begin{split} \label{e.phdyson}
D^r_\alpha &= [(d_\alpha^r)^{-1}-\Pi^r_\alpha]^{-1} \\
D^{\pm\mp}_\alpha & =(1+D^r_\alpha\Pi^r_\alpha)
d^{\pm\mp}_\alpha(1+\Pi^a_\alpha D^a_\alpha)
-D^r_\alpha\Pi^{\pm\mp}_\alpha D^a_\alpha. 
\end{split}\end{equation}
For self-energies, or ``polarizations'', we use the
second-order approximations
\begin{equation}\begin{split} \label{e.phself}
\Pi^{\pm\mp}_\alpha(\epsilon)
&=\iu\int\frac{\upd\omega_1}{2\pi}
\Tr[\gmat{\lambda}^\alpha\gmat{G}^{\pm\mp}(\omega_1)
\gmat{\lambda}^\alpha\gmat{G}^{\mp\pm}(\omega_1-\epsilon)] \\
\Pi^{r}_\alpha(\epsilon)
&=-\iu\int\frac{\upd\omega_1}{2\pi}
\Tr[\gmat{\lambda}^\alpha\gmat{G}^{\pm\mp}(\omega_1)
\gmat{\lambda}^\alpha\gmat{G}^{a}(\omega_1-\epsilon) \\
&+\gmat{\lambda}^\alpha\gmat{G}^{r}(\omega_1)\gmat{\lambda}^\alpha
\gmat{G}^{\pm\mp}(\omega_1-\epsilon)
] \\
\Pi^{a}_\alpha(\epsilon)
&=-\iu\int\frac{\upd\omega_1}{2\pi}
\Tr[\gmat{\lambda}^\alpha\gmat{G}^{\pm\mp}(\omega_1)
\gmat{\lambda}^\alpha\gmat{G}^{r}(\omega_1-\epsilon) \\
&+\gmat{\lambda}^\alpha\gmat{G}^{a}(\omega_1)\gmat{\lambda}^\alpha
\gmat{G}^{\pm\mp}(\omega_1-\epsilon)
],
\end{split}\end{equation}
where we have dropped some unimportant zero-frequency terms.
Again either the upper or the lower signs must be chosen.
Note that $\Pi^{\pm\mp}_\alpha$ are purely imaginary and
satisfy the symmetry $\Pi^{+-}_\alpha(-\epsilon)=\Pi^{-+}_\alpha(\epsilon)$.

Equations (\ref{e.phself}) close the system of equations, 
and we are done. However, from a physical point of view, it is interesting 
to develop the equations slightly further.
Using the symmetry $D_\alpha^r-D_\alpha^a=D^{-+}_\alpha-D^{+-}_\alpha$, one
finds that Eqs.\ (\ref{e.phdyson}) may be rewritten in the form
\begin{equation}\begin{split} \label{e.forced}
D^r_\alpha(\epsilon) 
&=\frac{2\epsilon_\alpha}{\epsilon^2-\epsilon_\alpha^2
+\iu\eta\epsilon-\eta^2/4-2\epsilon_\alpha\Pi^r_\alpha(\epsilon)} \\
D_\alpha^{+-}(\epsilon) & = -2\pi\iu N_\alpha(\epsilon)\rho_\alpha(\epsilon) \\
D_\alpha^{-+}(\epsilon) & = -2\pi\iu (N_\alpha(\epsilon)+1)\rho_\alpha(\epsilon), \\
\end{split}\end{equation}
where we define the phonon density of states
\begin{equation}\begin{split} \label{e.fullphdos}
\rho_\alpha(\epsilon) & =-\frac{1}{\pi}\im D^r_\alpha(\epsilon), \\ 
\end{split}\end{equation}
which satisfies
$\rho_\alpha(-\epsilon)=-\rho_\alpha(\epsilon)$.
The quantity $N_\alpha(\epsilon)$ is the energy distribution function
of the vibrational quanta. In equilibrium $N_\alpha(\epsilon)=n(\epsilon)$, 
the Bose distribution. 
In general
$N_\alpha(\epsilon)=n(\epsilon)+\delta N_\alpha(\epsilon)$, where
$\delta N_\alpha(\epsilon)$ is a voltage-dependent non-equilibrium 
correction. 

By reshuffling the Keldysh equations
one may write 
\begin{equation}\begin{split} \label{e.dpmmp}
D^{+-}_\alpha(\epsilon)&=-D^r_\alpha[\iu n\epsilon\eta/\epsilon_\alpha
+\Pi^{+-}_{\alpha}]D^a_\alpha \\
D^{-+}_\alpha(\epsilon)&=-D^r_\alpha[\iu (1+n)\epsilon\eta/\epsilon_\alpha
+\Pi^{-+}_{\alpha}]D^a_\alpha. \\
\end{split}\end{equation}
Since $|D_\alpha^r|^2=\im D_\alpha^r/(\im\Pi^r_\alpha(\epsilon)-\eta\epsilon/2\epsilon_\alpha)$,
comparing these with Eqs.\ (\ref{e.forced}) it is easy to 
obtain explicit expressions 
for the distribution function $N_\alpha$:
\begin{equation}\begin{split} \label{e.distrib}
N_\alpha(\epsilon)
&=-\frac{1}{2}\frac{\im\Pi^{+-}_\alpha(\epsilon)+
n(\epsilon)\eta\epsilon/\epsilon_\alpha}
{\im\Pi^r_\alpha(\epsilon)-\eta\epsilon/2\epsilon_\alpha} \\
&=-\frac{1}{2}\frac{\im\Pi^{-+}_\alpha(\epsilon)+
[1+n(\epsilon)]\eta\epsilon/\epsilon_\alpha}
{\im\Pi^r_\alpha(\epsilon)-\eta\epsilon/2\epsilon_\alpha} - 1.
\end{split}\end{equation}
To get to the last line, we used
$\Pi^r_\alpha-\Pi^a_\alpha=-(\Pi^{-+}_\alpha-\Pi^{+-}_\alpha)$.
Note that we have not made any approximations to get to this result.
Using Eqs.\ (\ref{e.phself}), one finds that 
the following symmetries are valid:
$N_\alpha(-\epsilon)=-[N_\alpha(\epsilon)+1]$
and 
$\delta N_\alpha(-\epsilon)=-\delta N_\alpha(\epsilon)$.

\section{Current conservation} \label{s.conser}

The inelastic parts of the current in Eqs.\ (\ref{e.curr})
look very asymmetric in their $L$ or $R$ indices. 
Nevertheless, in stationary state
the currents calculated at $L$ and $R$ boundaries should be equal:
$I^L=I^R$.
There also appears to be some confusion
as to what sort of approximations are needed 
for current conservation \cite{Frederiksen04a,Paulsson05}.
Here we outline a proof of this property for our approximation.
For $I_{el}$ it is obvious, so 
we only consider the inelastic current. 

Inserting the self-energies from Eq.\ (\ref{e.fullsigmas}) into the 
expression for $I^L_{inel}$ [Eq.\ (\ref{e.curr0})], one finds
\begin{widetext}
\begin{equation} \label{e.cons}
\begin{split}
I_{inel}^L=&\frac{2e}{\hbar}
\sum_\alpha \int\frac{\upd\epsilon}{2\pi}
\int\frac{\upd\omega_1}{2\pi} 2\pi\rho_\alpha(\omega_1)
\Big\{ \\
&\Tr[
\gmat{G}^a(\epsilon-\frac{\omega_1}{2})
\gmat{\lambda}^\alpha
\gmat{G}^a(\epsilon+\frac{\omega_1}{2})
\gmat{\Gamma}_L(\epsilon+\frac{\omega_1}{2})
\gmat{G}^r(\epsilon+\frac{\omega_1}{2})
\gmat{\lambda}^\alpha
\gmat{G}^r(\epsilon-\frac{\omega_1}{2})
\gmat{\Gamma}_L(\epsilon-\frac{\omega_1}{2})
] \\
&\times [f_L(\epsilon+\frac{\omega_1}{2})
(N_\alpha(\omega_1)+1)-(f_L(\epsilon+\frac{\omega_1}{2})
+N_\alpha(\omega_1))f_L(\epsilon-\frac{\omega_1}{2})] \\
+&\Tr[
\gmat{G}^a(\epsilon-\frac{\omega_1}{2})
\gmat{\lambda}^\alpha
\gmat{G}^a(\epsilon+\frac{\omega_1}{2})
\gmat{\Gamma}_L(\epsilon+\frac{\omega_1}{2})
\gmat{G}^r(\epsilon+\frac{\omega_1}{2})
\gmat{\lambda}^\alpha
\gmat{G}^r(\epsilon-\frac{\omega_1}{2})
\gmat{\Gamma}_R(\epsilon-\frac{\omega_1}{2})
] \\
&\times [f_L(\epsilon+\frac{\omega_1}{2})
(N_\alpha(\omega_1)+1)-(f_L(\epsilon+\frac{\omega_1}{2})
+N_\alpha(\omega_1))f_R(\epsilon-\frac{\omega_1}{2})] 
\Big\}.
\end{split}
\end{equation}
\end{widetext}
Here the first term may be shown to vanish as follows. 
Since Fermi and Bose functions $f$ and $n$ satisfy
$f(x)[n(y)+1]-[f(x)+n(y)]f(x-y)=0$, we have
\begin{equation}\begin{split} \label{e.finite}
f_L(x)&[N_\alpha(y)+1]-[f_L(x)+N_\alpha(y)]f_L(x-y) \\
&=\delta N_\alpha(y)[f_L(x)-f_L(x-y)].
\end{split}\end{equation}
Now, noting that 
$\delta N_\alpha(-\epsilon)=-\delta N_\alpha(\epsilon)$, and
$\rho_\alpha(-\epsilon)=-\rho_\alpha(\epsilon)$, 
and additionally assuming that all matrices under the trace 
are symmetric ($\gmat{G}^{rT}=\gmat{G}^r$, 
$[\gmat{\Gamma}_L]^T=\gmat{\Gamma}_L$ etc.)
it is seen that the
integrand is odd and thus the energy integral vanishes. 
This is true also if $\delta N_\alpha(\epsilon)$ has a $\sim 1/\epsilon$ 
divergence 
at $\epsilon=0$, since the product in Eq.\ 
(\ref{e.finite}) remains finite.

The second term in Eq.\ (\ref{e.cons}) 
is clearly symmetric upon interchanging
$L$ and $R$, and changing the overall sign.
Note that the proof does not rely on things like a mirror symmetry 
of the geometry, and remains unchanged if we replace 
$\gmat{G}^r\rightarrow\gmatx{G}^r$.
Thus our expressions are always  
``current conserving'', as defined by the condition 
$I^L=I^R$.

\section{General perturbative current formulas} \label{s.form}

Here we shall write down the perturbative current 
formulas of Eqs.\ (\ref{e.pertcurr}) in 
a slightly different form, which will make the
so-called wide-band approximation 
more transparent. However, here we shall be making no 
approximations in addition to the second-order perturbation theory 
which we have already introduced.
Nevertheless, it must be stressed that, due to the 
second-order approximation, all higher-order terms
in the following expressions are strictly speaking not
warranted. This concerns the approximations
made for the vibrational density of states $\rho_{\alpha}$
and the distribution $N_\alpha$, which should in principle 
both be of zeroth order in the electron-vibration coupling constant.
In the approximations which we use for $N_\alpha$, 
this is not necessarily the case. 
Thus our approach is not purely lowest-oder perturbation theory.
However, the self-consistent Born scheme (SCBA)
is not really any better is this respect.

The inelastic current $I_{in}$,
and the elastic parts $I_{el}^0$ and $\delta I_{el}$ 
are obtained from Eqs.\ (\ref{e.pertcurr}).
Inserting the self-energies [Eqs.\ (\ref{e.fullsigmas})] into these
formulas, they may be rewritten as follows:
\begin{widetext}
\begin{equation}\begin{split}
I^0_{el} & = \frac{2e}{h}\int\upd\epsilon
T_0(\epsilon)[f_L(\epsilon)-f_R(\epsilon)] \\
\end{split}\end{equation}
\begin{equation}\begin{split}
\delta I_{el} & = \frac{2e}{h}
\int\upd\epsilon 
\sum_\alpha\sum_{\sigma=\pm1}
\sigma
\int_0^{\infty}\upd\omega_1\rho_\alpha(\omega_1)
\Big\{
T^{ec}_{\sigma\alpha}(\epsilon,\omega_1)N_\alpha(\sigma\omega_1) 
+T^{ecL}_{\sigma\alpha}(\epsilon,\omega_1)f_L(\epsilon_{\sigma\alpha})
+T^{ecR}_{\sigma\alpha}(\epsilon,\omega_1)f_R(\epsilon_{\sigma\alpha})
\Big\} \\
&\times[f_L(\epsilon)-f_R(\epsilon)]
+\frac{2e}{h}\int\upd\epsilon\sum_\alpha \Big\{
-J^L_\alpha(\epsilon)-J^R_\alpha(\epsilon) 
+
T^{II}_\alpha(\epsilon)[J_\alpha^{IIL}+J_\alpha^{IIR}]
\Big\}
[f_L(\epsilon)-f_R(\epsilon)] \\
\end{split}\end{equation}
\begin{equation}\begin{split}
I_{inel}^L & = \frac{2e}{h}\int\upd\epsilon \sum_\alpha\sum_{\sigma=\pm1}
\sigma 
\int_0^{\infty}\upd\omega_1\rho_\alpha(\omega_1)
T^{in}_{\sigma\alpha}(\epsilon,\omega_1)
\Big\{ 
N_\alpha(\sigma\omega_1)
f_L(\epsilon)[1-f_R(\epsilon_{\sigma\alpha})] \\
&+N_\alpha(-\sigma\omega_1)
f_R(\epsilon_{\sigma\alpha})[1-f_L(\epsilon)]
\Big\},
\end{split}\end{equation}
where $\epsilon_{\sigma\alpha}=\epsilon+\sigma\omega_1$.
The elastic transmission is given by
\begin{equation}\begin{split}
T_0(\epsilon) = \Tr[\gmatx{G}^r(\epsilon)\gmat{\Gamma}_R(\epsilon)\gmatx{G}^a(\epsilon)
\gmat{\Gamma}_L(\epsilon)]
\end{split}\end{equation}
and the inelastic prefactor by
\begin{equation}\begin{split}
T^{in}_{\sigma\alpha}(\epsilon,\omega_1) = \Tr[\gmatx{G}^r(\epsilon_{\sigma\alpha})
\gmat{\Gamma}_R(\epsilon_{\sigma\alpha})\gmatx{G}^a(\epsilon_{\sigma\alpha})
\gmat{\lambda}^\alpha
\gmatx{G}^a(\epsilon)\gmat{\Gamma}_L(\epsilon)\gmatx{G}^r(\epsilon)
\gmat{\lambda}^\alpha],
\end{split}\end{equation}
while the factors in the elastic correction are
\begin{equation}\begin{split} \label{e.eccoeffs}
T^{ec}_{\sigma\alpha}(\epsilon,\omega_1)&=2\re\Tr[
\gmatx{G}^r(\epsilon)\gmat{\Gamma}_R(\epsilon)\gmatx{G}^a(\epsilon)
\gmat{\Gamma}_L(\epsilon)\gmatx{G}^r(\epsilon)\gmat{\lambda}^\alpha
\gmatx{G}^r(\epsilon_{\sigma\alpha})\gmat{\lambda}^\alpha] \\
T^{ecL}_{\sigma\alpha}(\epsilon,\omega_1)&=\im\Tr[
\gmatx{G}^r(\epsilon)\gmat{\Gamma}_R(\epsilon)\gmatx{G}^a(\epsilon)
\gmat{\Gamma}_L(\epsilon)\gmatx{G}^r(\epsilon)\gmat{\lambda}^\alpha
\gmatx{G}^r(\epsilon_{\sigma\alpha})\gmat{\Gamma}_L(\epsilon_{\sigma\alpha})
\gmatx{G}^a(\epsilon_{\sigma\alpha})
\gmat{\lambda}^\alpha] \\
T^{ecR}_{\sigma\alpha}(\epsilon,\omega_1)&=\im\Tr[
\gmatx{G}^r(\epsilon)\gmat{\Gamma}_R(\epsilon)\gmatx{G}^a(\epsilon)
\gmat{\Gamma}_L(\epsilon)\gmatx{G}^r(\epsilon)\gmat{\lambda}^\alpha
\gmatx{G}^r(\epsilon_{\sigma\alpha})\gmat{\Gamma}_R(\epsilon_{\sigma\alpha})
\gmatx{G}^a(\epsilon_{\sigma\alpha})
\gmat{\lambda}^\alpha]. \\
\end{split}\end{equation}
The most complicated part are the integrals
\begin{equation}\begin{split}\label{e.nasty1}
J^L_\alpha(\epsilon)&=
\int\frac{\upd\omega_1}{2\pi}
2\re[D^r_\alpha(\omega_1)]
\re\Tr[
\gmatx{G}^r(\epsilon)\gmat{\Gamma}_R(\epsilon)\gmatx{G}^a(\epsilon)
\gmat{\Gamma}_L(\epsilon)\gmatx{G}^r(\epsilon)
\gmat{\lambda}^\alpha
\gmatx{G}^r(\epsilon-\omega_1)
\gmat{\Gamma}_L(\epsilon-\omega_1)
\gmatx{G}^a(\epsilon-\omega_1)
\gmat{\lambda}^\alpha]f_L(\epsilon-\omega_1) \\
J^R_\alpha(\epsilon)&=
\int\frac{\upd\omega_1}{2\pi}
2\re[D^r_\alpha(\omega_1)]
\re\Tr[
\gmatx{G}^r(\epsilon)\gmat{\Gamma}_R(\epsilon)\gmatx{G}^a(\epsilon) 
\gmat{\Gamma}_L(\epsilon)\gmatx{G}^r(\epsilon)
\gmat{\lambda}^\alpha
\gmatx{G}^r(\epsilon-\omega_1)
\gmat{\Gamma}_R(\epsilon-\omega_1)
\gmatx{G}^a(\epsilon-\omega_1)
\gmat{\lambda}^\alpha]f_R(\epsilon-\omega_1)
\end{split}\end{equation}
and the coefficients coming from the last term in Eq.\ (\ref{e.fullsigmas}) 
for $\gmat{\Sigma}^r$ 
\begin{equation}\begin{split} \label{e.nasty2}
T^{II}_\alpha(\epsilon)&=2\re\Tr[
\gmatx{G}^r(\epsilon)\gmat{\Gamma}_R(\epsilon)\gmatx{G}^a(\epsilon)
\gmat{\Gamma}_L(\epsilon)\gmatx{G}^r(\epsilon)\gmat{\lambda}^\alpha] \\
J_\alpha^{IIL}&=D^r_\alpha(0)\int\frac{\upd\omega_1}{2\pi}
\Tr[
\gmatx{G}^r(\omega_1)\gmat{\Gamma}_L(\omega_1)\gmatx{G}^a(\omega_1)
\gmat{\lambda}^\alpha]f_L(\omega_1) \\
J_\alpha^{IIR}&=D^r_\alpha(0)\int\frac{\upd\omega_1}{2\pi}
\Tr[
\gmatx{G}^r(\omega_1)\gmat{\Gamma}_R(\omega_1)\gmatx{G}^a(\omega_1)
\gmat{\lambda}^\alpha]f_R(\omega_1).
\end{split}\end{equation}
\end{widetext}

In the equations above, the phonon density of states
$\rho_\alpha(\epsilon)$ includes all possible broadening
effects and shifts of the bare vibrational frequencies.
Without these effects 
$\rho_\alpha(\epsilon)=\delta(\epsilon-\hbar\omega_\alpha)-\delta(\epsilon+\hbar\omega_\alpha)$, and Eqs.\ (\ref{e.nasty1}) must be evaluated as principal
part integrals.

In the so-called wide band limit, discussed in the text,
the $\epsilon$ and $\omega_1$
dependences of the coefficients $T_0$, $T^{in}_\alpha$,
$T^{ec}_\alpha$ and $T^{ecL,R}_\alpha$, 
may be dropped and they may be simply evaluated at the Fermi energy. 
Then the integrals over products of Fermi functions in the corresponding 
current terms can be done analytically.
One also finds that all terms of $\delta I_{el}$ which do not involve
Eqs.\ (\ref{e.nasty1}) or (\ref{e.nasty2}) yield conductance contributions
$\delta G_{el}(\epsilon)=\upd \delta I_{el}/\upd V$
which are symmetric in the bias: 
$\delta G_{el}(-V)=\delta G_{el}(V)$.
We call the sum of these terms $\delta I^{sym}_{el}$,
and $I^{sym}=I^0_{el}+\delta I^{sym}_{el}+I_{inel}$.
The same may be done to the trace expressions in Eqs.\ (\ref{e.nasty1}),
and one finds that the corresponding part in the current $\delta I_{el}$
yields the asymmetric current $I^{asy}$ \cite{Paulsson05}.

The contribution of Eqs.\ (\ref{e.nasty2})
to the current is typically very small in the small-voltage limit 
which we are considering. 
Furthermore, since they do not introduce any relation between 
the voltage and the vibrational frequencies, 
they cannot give a contribution to the conductance steps.
Therefore, we drop them for the sake of simplicity.


\end{document}